\documentclass{article}
\usepackage{amssymb}

\input{tcilatex}

\begin{document}

\begin{center}
{\large APPLICATION OF LIE TRANSFORMATION GROUP METHODS TO\smallskip }

{\large CLASSICAL THEORIES OF PLATES AND RODS}

\bigskip

V. VASSILEV\footnote{%
E-mail: vassil@bgcict.acad.bg} and P. DJONDJOROV\footnote{%
E-mail: padjon@bgcict.acad.bg}

{\small Institute of Mechanics, Bulgarian Academy of Sciences}

{\small Acad. G. Bontchev St., Block 4, 1113 Sofia, Bulgaria\bigskip }

\bigskip
\end{center}

\begin{quotation}
\noindent \textbf{Abstract---}In the present paper, a class of partial
differential equations related to various plate and rod problems is studied
by Lie transformation group methods. A system of equations determining the
generators of the admitted point Lie groups (symmetries) is derived. A
general statement of the associated group-classification problem is given. A
simple intrinsic relation is deduced allowing to recognize easily the
variational symmetries among the ''ordinary'' symmetries of a self-adjoint
equation of the class examined. Explicit formulae for the conserved currents
of the corresponding (via Noether's theorem) conservation laws are
suggested. Solutions of group-classification problems are presented for
subclasses of equations of the foregoing type governing stability and
vibration of plates, rods and fluid conveying pipes resting on variable
elastic foundations and compressed by axial forces. The obtained
group-classification results are used to derive conservation laws and
group-invariant solutions readily applicable in plate statics or rod
dynamics.\bigskip
\end{quotation}

\begin{center}
\medskip

\medskip 1. INTRODUCTION
\end{center}

A wide variety of classical theories of plates and rods\footnote{%
In this work, following Antman (1984) we use ''rod'' as a generic name for
''arch'', ''bar'', ''beam'', ''ring'', ''column'', ''tube'', ''pipe'', etc.
We employ ''rod'' in the intuitive sense of a slender solid body.} rest on
linear fourth-order partial differential equations in one dependent and two
independent variables. Some of them, such as the Poisson-Kirchhoff type
theories for small bending of plates, are developed within the framework of
the linear elastostatics to determine the state of equilibrium of thin
elastic plates in terms of the transversal displacement of the plate middle
plane, the derived governing equations providing the background for solving
problems concerning stability and vibration of such structural elements.
Others (among them -- the dynamic theory of Bernoulli-Euler beams, for
instance) are deduced on the ground of the linear elastodynamics to describe
the dynamic behaviour of rods in terms of the transversal displacement of
the rod axis.

The aim of the present paper is to study the invariance properties
(symmetries) of the equations of the foregoing type with respect to local
Lie groups of point transformations of the involved independent and
dependent variables. The work is motivated both by the aforesaid wide
applicability of the equations in question in structural mechanics, and by
the remarkable efficiency demonstrated by the symmetry methods, especially
when applied to differential equations arising in physics and engineering.

Actually, once the invariance properties of a given differential equation
are established, several important applications of its symmetries arise.
First, it is possible to distinguish classes of solutions to this equation
invariant under the transformations of symmetry groups admitted. The
determination of such a group-invariant solution assumes solving a reduced
equation involving less independent variables than the original one. Typical
examples of group-invariant solutions are axisymmetric solutions,
self-similar solutions, travelling waves, etc., which have proved to be
quite useful in many branches of physics and engineering. For a self-adjoint
differential equation another substantial application of its symmetries is
available. As is well known, the self-adjoint equations are the
Euler-Lagrange equations of a certain action functional. If a one-parameter
symmetry group of such an equation turned out to be its variational symmetry
as well, that is a symmetry of the associated action functional, then
Noether's theorem guarantees the existence of a conservation law for the
smooth solutions of this equation. Needless to recall or discuss here the
fundamental role that the conserved quantities and conservation laws (or the
corresponding integral relations, i.e. the balance laws) have played in
natural sciences, but it is worthy to point out that the available
conservation laws (balance laws) should not be overlooked (as it is often
done) in the examination of discontinuous solutions (acceleration waves,
shock waves, etc.) or in the numerical analysis (when constructing finite
difference schemes or verifying numerical results, for instance) of any
system of differential equations of physical interest. It should be remarked
also that the path-independent integrals (such as the well known $J$-, $L$-
and $M$-integrals) related to the conservation laws are basic tools in
fracture analysis of solids and structures.

The aforementioned and many other applications of the symmetries of
differential equations and variational problems as well as the foundations
of the Lie transformation group methods, including the basic notions,
statements and techniques, can be found in the books by Ovsiannikov (1982),
Ibragimov (1985), Bluman and Kumei (1989) and Olver (1993) (see also the
references given in these books). In the present paper, however, as fare as
the application of the symmetry groups of the equations studied is
concerned, our attention is restricted to the constructing of
group-invariant solutions and conservation laws. Of course, the first task
is to find these symmetry groups, and as here we do not deal with a single
differential equation but with a class of differential equations, this means
to solve a group-classification problem.

The layout of the paper is as follows. A detailed description of the
differential equations to be studied as well as the variational statement
for the self-adjoint equations among them are given in Section 2. Several
examples of mechanical systems governed by such equations complete this
Section. In Section 3, a system of equations determining the generators of
the symmetry groups admitted by the equations of the class considered is
derived, the invariance properties inherent to the whole class because of
its linearity and homogeneity being taken into account, and then the general
statement of the associated group-classification problem is given. After
that, the variational symmetries of the self-adjoint equations of the
examined class are investigated. A simple intrinsic relation allowing to
recognize easily the variational symmetries among the ''ordinary'' point Lie
symmetries of such an equation is deduced, and explicit formulae for the
conserved currents of the conservation laws corresponding to the variational
symmetries via Noether's theorem are suggested. Group-classification
results, conservation laws and group-invariant solutions are presented in
Section 4 for differential equations governing stability and vibration of
plates of Poisson-Kirchhoff type. Similar results are displayed in Section 5
for equations governing vibration of rods on a variable elastic foundation
and dynamic stability of fluid conveying pipes or rods compressed by axial
forces. In the reminder of this Section, conservation laws for rods derived
in the present contribution are compared to other ones reported in the
literature.

\begin{center}
\medskip

\medskip 2.BASIC EQUATIONS
\end{center}

Consider a fourth-order homogeneous linear partial differential equation 
\begin{equation}
A^{\alpha \beta \gamma \delta }(x)w_{\alpha \beta \gamma \delta }+A^{\alpha
\beta \gamma }(x)w_{\alpha \beta \gamma }+A^{\alpha \beta }(x)w_{\alpha
\beta }+A^{\alpha }(x)w_{\alpha }+A(x)w=0,  \label{BasicEqn}
\end{equation}
in two independent variables $x=(x^{1},x^{2})$ and one dependent variable $%
w(x)$. Here and throughout: Greek indices have the range 1, 2, and the usual
summation convention over a repeated index (one subscript and one
superscript) is employed; $w_{\alpha _{1}\alpha _{2}...\alpha _{k}}$ $%
(k=1,2,...)$ denote the $k$-th order partial derivatives of the dependent
variable, i.e. 
\[
w_{\alpha _{1}\alpha _{2}...\alpha _{k}}=\frac{\partial \,^{k}w}{\partial
\/x^{\alpha _{1}}\partial \/x^{\alpha _{2}}...\partial \/x^{\alpha _{k}}}%
\quad (k=1,2,...). 
\]
Further, a similar notation will be used for the partial derivatives of any
other function of the variables $x^{1},x^{2}$ but, in this case, the indices
indicating the differentiation will be preceded by a coma, e.g. 
\[
A_{,\alpha _{1}\alpha _{2}...\alpha _{k}}^{\alpha \beta \gamma \delta }{=}%
\frac{\partial \,^{k}A^{\alpha \beta \gamma \delta }}{\partial \/x^{\alpha
_{1}}\partial \/x^{\alpha _{2}}...\partial \/x^{\alpha _{k}}}\quad
(k=1,2,...). 
\]
The coefficients of equation (\ref{BasicEqn}) are supposed to be smooth
functions possessing as many derivatives as may be required on a certain
domain of interest, and to be symmetric under any permutation of their
indices, i.e. 
\[
A^{\alpha \beta \gamma \delta }=A^{\beta \gamma \delta \alpha }=A^{\gamma
\delta \alpha \beta }=A^{\delta \alpha \beta \gamma },\quad A^{\alpha \beta
\gamma }=A^{\beta \gamma \alpha }=A^{\gamma \alpha \beta },\quad A^{\alpha
\beta }=A^{\beta \alpha }. 
\]
Using the total derivative operators 
\[
D_{\alpha }=\frac{\partial }{\partial \,x^{\alpha }}+w_{\alpha }\frac{%
\partial }{\partial \,w}+w_{\alpha \mu }\frac{\partial }{\partial \,w_{\mu }}%
+w_{\alpha \mu \nu }\frac{\partial }{\partial \,w_{\mu \nu }}+w_{\alpha \mu
\nu \sigma }\frac{\partial }{\partial \,w_{\mu \nu \sigma }}+\cdots , 
\]
the equation (\ref{BasicEqn}) may be written in the form 
\begin{equation}
\mathcal{D}[w]=0,  \label{BasicEqn1}
\end{equation}
where $\mathcal{D}$ is the linear differential operator given by the
expression 
\begin{equation}
\mathcal{D}=A^{\alpha \beta \gamma \delta }D_{\alpha }D_{\beta }D_{\gamma
}D_{\delta }+A^{\alpha \beta \gamma }D_{\alpha }D_{\beta }D_{\gamma
}+A^{\alpha \beta }D_{\alpha }D_{\beta }+A^{\alpha }D_{\alpha }+A.
\label{DO}
\end{equation}

An equation of form (\ref{BasicEqn1}) is the Euler-Lagrange equation
associated with a certain variational problem involving only one dependent
variable if and only if the differential operator $\mathcal{D}$ is
self-adjoint, that is 
\begin{equation}
\mathcal{D}=\mathcal{D}^{\ast },  \label{S-A-1}
\end{equation}
where $\mathcal{D}^{\mathcal{\ast }}$ is the (formal) adjoint operator of $%
\mathcal{D}$ (cf. Olver, 1993). The explicit form of $\mathcal{D}^{\mathcal{%
\ast }}$ is 
\begin{equation}
\mathcal{D}^{\mathcal{\ast }}=D_{\alpha }D_{\beta }D_{\gamma }D_{\delta
}A^{\alpha \beta \gamma \delta }-D_{\alpha }D_{\beta }D_{\gamma }A^{\alpha
\beta \gamma }+D_{\alpha }D_{\beta }A^{\alpha \beta }-D_{\alpha }A^{\alpha
}+A.  \label{ADifO-1}
\end{equation}
In such a case, (\ref{BasicEqn1}) can be associated with the variational
problem for the functional 
\[
A[w]=\int \frac{1}{2}w\mathcal{D}[w]dx^{1}dx^{2}, 
\]
since the application of the Euler operator 
\[
\mathsf{E}=\frac{\partial }{\partial \,w}-D_{\mu }\frac{\partial }{\partial
\,w_{\mu }}+D_{\mu }D_{\nu }\frac{\partial }{\partial \,w_{\mu \nu }}-D_{\mu
}D_{\nu }D_{\sigma }\frac{\partial }{\partial \,w_{\mu \nu \sigma }}+D_{\mu
}D_{\nu }D_{\sigma }D_{\tau }\frac{\partial }{\partial \,w_{\mu \nu \sigma
\tau }}-\cdots 
\]
on the Lagrangian density 
\begin{equation}
L=\frac{1}{2}w\mathcal{D}[w],  \label{LD-1}
\end{equation}
yields 
\begin{equation}
\mathcal{D}[w]=\mathsf{E}(L),  \label{E-L-1}
\end{equation}
due to (\ref{S-A-1}) and (\ref{ADifO-1}).

Let us give several examples of plate and rod structures whose governing
equations are self-adjoint and belong to the class specified above.
Henceforward, when regarding plates the variables $x^{1},x^{2}$ will
represent the coordinates of the plate middle plane. As for the rod
problems, $x^{1}$ will be associated with the spatial variable along the rod
axis, and $x^{2}$ -- with the time. In both cases, the dependent variable $w$
will represent the transversal displacement field.

\textbf{Example 1}. \textit{Small bending of plates resting on elastic
foundations.} Consider a thin elastic plate of variable bending rigidity $%
D(x)$ resting on an elastic foundation of Winkler type with variable modulus 
$k(x)$ and subjected to an edge loading leading to the appearance of
nonuniform membrane stresses $N^{\alpha \beta }(x)$. In this physical
situation, the equation governing the small bending of the plate assumes the
following form 
\begin{equation}
\Delta \lbrack D\Delta w]-[(1-\nu )\varepsilon ^{\alpha \mu }\varepsilon
^{\beta \nu }D_{,\alpha \beta }-N^{\mu \nu }]w_{\mu \nu }+kw=0\,,
\label{Plate-Eqn}
\end{equation}
the membrane stress tensor $N^{\alpha \beta }$ being symmetric, $N^{\alpha
\beta }=N^{\beta \alpha }$, and divergence free, i.e. $N_{,\mu }^{\alpha \mu
}=0$. Here: $\nu $ is Poisson's ratio; $\Delta $ is the Laplace operator,
that is $\Delta \equiv \delta ^{\alpha \beta }\partial ^{2}/\partial
\,x^{\alpha }\partial \,x^{\beta }$, where $\delta ^{\alpha \beta }$ is the
Kronecker delta symbol ($\delta ^{11}=\delta ^{22}=1,\;\delta ^{12}=\delta
^{21}=0$) and $\varepsilon ^{\alpha \beta }$ is the alternating symbol ($%
\varepsilon ^{11}=\varepsilon ^{22}=0,\;\varepsilon ^{12}=-\varepsilon
^{21}=1$).

\textbf{Example 2}\textit{. Elastodynamics of Bernoulli-Euler beams. }%
Consider a nonhomogeneous Bernoulli-Euler beam with bending rigidity $%
B(x^{1})$ and inertia term $H(x^{1})$. The differential equation governing
the small vibration of such a beam is (Chien at al., 1993): 
\[
Bw_{1111}+2B_{,1}w_{111}+B_{,11}w_{11}+Hw_{22}=0. 
\]

\textbf{Example 3.}\textit{\ Elastic beams resting on elastic foundations.}
Consider an elastic beam of constant bending rigidity $K$ and constant mass
density $m$\ (mass of the beam per unit length), resting on an elastic
foundation with variable modulus $k\/(x)$. Suppose it is subjected to a
constant follower force $p$. Then, according to the Bernoulli-Euler theory,
the differential equation for small transverse vibration of the beam is (see
Smith and Herrmann, 1972): 
\begin{equation}
Kw_{1111}+pw_{11}+k(x)w+mw_{22}=0\,.  \label{eq2.12}
\end{equation}

\textbf{Example 4.} \textit{Pipes conveying fluid.} Consider an elastic
circular-cylindrical pipe of uniform outer radius of the pipe cross section,
which is supposed to be small in comparison with certain characteristic pipe
length (for a simply supported pipe say the length of the span). Let the
pipe conveys inviscid incompressible fluid with a flow velocity $U=const$.
Then, the equation of motion is (see Gregory and Paidoussis, 1966): 
\begin{equation}
EJw_{1111}+MU^{2}w_{11}+2MUw_{12}+(m+M)w_{22}=0\,,  \label{eq2.16}
\end{equation}
where $E$ is Young's modulus of the pipe material, $J$ is the (axial) moment
of inertia of the pipe cross section, $m$ is the mass (constant) of the pipe
per unit length, $M$ is the mass (also constant) of the fluid per unit
length.

Combining and generalizing equations (\ref{eq2.12}) and (\ref{eq2.16})
presented in Examples 3 and 4, in Section 5 we will pay particular attention
to the differential equations of the form 
\begin{equation}
\gamma w_{1111}+\chi ^{\alpha \beta }w_{\alpha \beta }+\kappa (x)w=0\,,
\label{eq2.21}
\end{equation}
where $\gamma \neq 0$ and $\chi ^{\alpha \beta }$ are real constants, while $%
\kappa (x)$ is a smooth function.

\medskip

\begin{center}
\medskip 3. SYMMETRIES AND CONSERVATION LAWS
\end{center}

Consider a local one-parameter Lie group of point transformations acting on
some open subset $\Omega $ of the space $\mathbf{R}^{3}$ representing the
independent and dependent variables $x^{1},x^{2},w$ involved in our basic
equation (\ref{BasicEqn1}). The infinitesimal generator of such a group is a
vector field $X$ on $\mathbf{R}^{3}$, 
\begin{equation}
X=\xi ^{\mu }\left( x,w\right) \frac{\partial }{\partial x^{\mu }}+\eta
\left( x,w\right) \frac{\partial }{\partial w},  \label{eq3.2}
\end{equation}
whose components $\xi ^{\mu }(x,w)$ and $\eta $$(x,w)$ are supposed to be
functions of class $C^{\infty }$on $\Omega $. By virtue of Theorem 2.31
(Olver, 1993), a vector field $X$ of form (\ref{eq3.2}) generates a point
Lie symmetry group of equation (\ref{BasicEqn1}) if and only if there exists
a function $\lambda $ depending on $x$, $w$ and derivatives of $w$ (that is
a differential function) such that the following infinitesimal criterion of
invariance, 
\begin{equation}
\mathrel{\mathop{X}\limits_{4}}\left( \mathcal{D}[w]\right) -\lambda 
\mathcal{D}[w]=0,  \label{eq3.3}
\end{equation}
holds; here $\mathrel{\mathop{X}\limits_{k}}$ denote the $k$\textrm{-th}
prolongation of $X$ (Ovsiannikov, 1982).\pagebreak

The invariance criterion (\ref{eq3.3}) leads, through the standard
computational procedure (see, e.g. Ovsiannikov, 1982 or Olver, 1993), to the
following results:

(i) each equation of form (\ref{BasicEqn1}) being linear and homogeneous is
invariant under the point Lie groups generated by the vector fields 
\begin{equation}
X_0=w\frac \partial {\partial w},\quad X_u=u\left( x\right) \frac \partial
{\partial w},  \label{Kernel}
\end{equation}
where $u\left( x\right) $ is an arbitrary solution of the equation
considered, the invariance criterion (\ref{eq3.3}) being fulfilled with $%
\lambda =1$ for $X_0$, and $\lambda =0$ for the generators $X_u$;

(ii)\textrm{\ }an equation of form (\ref{BasicEqn1}) admits other vector
fields (\ref{eq3.2}), in addition to the aforementioned (\ref{Kernel}), if
and only if they have the special form 
\begin{equation}
X=\xi ^{\mu }\left( x\right) \frac{\partial }{\partial x^{\mu }}+\sigma
\left( x\right) w\frac{\partial }{\partial w},  \label{eq3.9}
\end{equation}
the functions $\xi ^{\mu }\left( x\right) $ and $\sigma \left( x\right) $
being nontrivial solutions of the following system of determining equations
(called further the DE system for easy reference):\smallskip\ 
\begin{equation}
\qquad \;\xi ^{\mu }A_{,\mu }^{\alpha \beta \gamma \delta }+(\sigma -\lambda
)A^{\alpha \beta \gamma \delta }-A^{\alpha \beta \gamma \mu }\xi _{,\mu
}^{\delta }-A^{\alpha \beta \mu \delta }\xi _{,\mu }^{\gamma }-A^{\alpha \mu
\gamma \delta }\xi _{,\mu }^{\beta }-A^{\mu \beta \gamma \delta }\xi _{,\mu
}^{\alpha }=0,  \label{eq3.14}
\end{equation}
\begin{equation}
\begin{tabular}{l}
$4A^{\alpha \beta \gamma \mu }\sigma _{,\mu }-2A^{\alpha \beta \mu \nu }\xi
_{,\mu \nu }^{\gamma }-2A^{\alpha \gamma \mu \nu }\xi _{,\mu \nu }^{\beta
}-2A^{\beta \gamma \mu \nu }\xi _{,\mu \nu }^{\alpha }\smallskip $ \\ 
$\quad \quad \ \,\quad \quad \quad \;\quad +\xi ^{\mu }A_{,\mu }^{\alpha
\beta \gamma }+(\sigma -\lambda )A^{\alpha \beta \gamma }-A^{\alpha \beta
\mu }\xi _{,\mu }^{\gamma }-A^{\alpha \mu \gamma }\xi _{,\mu }^{\beta
}-A^{\mu \beta \gamma }\xi _{,\mu }^{\alpha }=0,$%
\end{tabular}
\label{eq3.15}
\end{equation}
\begin{equation}
\begin{tabular}{l}
$6A^{\alpha \beta \mu \nu }\sigma _{,\mu \nu }-2A^{\alpha \mu \nu \sigma
}\xi _{,\mu \nu \sigma }^{\beta }-2A^{\beta \mu \nu \sigma }\xi _{,\mu \nu
\sigma }^{\alpha }\smallskip $ \\ 
$\quad \quad \quad \quad \quad \quad \quad \;\quad +3A^{\alpha \beta \mu
}\sigma _{,\mu }-(3/2)A^{\alpha \mu \nu }\xi _{,\mu \nu }^{\beta
}-(3/2)A^{\beta \mu \nu }\xi _{,\mu \nu }^{\alpha }\smallskip $ \\ 
$\quad \quad \quad \quad \quad \quad \quad \quad \quad \quad \quad \quad
\;\quad +\xi ^{\mu }A_{,\mu }^{\alpha \beta }+(\sigma -\lambda )A^{\alpha
\beta }-A^{\alpha \mu }\xi _{,\mu }^{\beta }-A^{\mu \beta }\xi _{,\mu
}^{\alpha }=0,$%
\end{tabular}
\label{eq3.16}
\end{equation}
\begin{equation}
\begin{tabular}{l}
$4A^{\alpha \mu \nu \sigma }\sigma _{,\mu \nu \sigma }-A^{\mu \nu \sigma
\tau }\xi _{,\mu \nu \sigma \tau }^{\alpha }\smallskip $ \\ 
$\quad \quad \quad \quad \quad \ \quad \quad \quad +3A^{\alpha \mu \nu
}\sigma _{,\mu \nu }-A^{\mu \nu \sigma }\xi _{,\mu \nu \sigma }^{\alpha
}\smallskip $ \\ 
$\quad \quad \quad \ \quad \quad \quad \quad \quad \ \ \quad \quad \ \quad
\quad \quad +2A^{\alpha \mu }\sigma _{,\mu }-A^{\mu \nu }\xi _{,\mu \nu
}^{\alpha }\smallskip $ \\ 
$\quad \quad \quad \ \quad \quad \quad \quad \ \ \quad \quad \quad \quad
\quad \ \quad \quad \quad \quad \;\;\quad +\xi ^{\mu }A_{,\mu }^{\alpha
}+(\sigma -\lambda )A^{\alpha }-A^{\mu }\xi _{,\mu }^{\alpha }=0,$%
\end{tabular}
\label{eq3.17}
\end{equation}
\begin{equation}
\quad \quad \quad \quad \quad A^{\alpha \beta \gamma \delta }\sigma
_{,\alpha \beta \gamma \delta }+A^{\alpha \beta \gamma }\sigma _{,\alpha
\beta \gamma }+A^{\alpha \beta }\sigma _{,\alpha \beta }+A^{\alpha }\sigma
_{,\alpha }+\xi ^{\mu }A_{,\mu }+(\sigma -\lambda )A=0,\smallskip
\label{eq3.18}
\end{equation}
for a certain function $\lambda $ depending on $x^{1}$ and $x^{2}$ only.
(Here, by a trivial solution we mean not only $\xi ^{\mu }=0,\,\sigma =0,$
but also $\xi ^{\mu }=0,\,\sigma =c=const\neq 0$, since the latter leads to
the vector field $cX_{0}$ generating the same group as $X_{0}$ which is
already identified to be admitted by each equation of the type considered.)

Thus, given an equation of form (\ref{BasicEqn1}), the question is whether
there exist vector fields $X\neq cX_{0}$ of form (\ref{eq3.9}) which leave
it invariant, and the answer depends on whether the respective DE system has
at least one nontrivial solution. In this context the coefficients of the
equation are supposed to be known functions, and thereby (\ref{eq3.14}) -- (%
\ref{eq3.18}) constitute an over-determined system of linear homogeneous
partial differential equations with respect to the unknowns $\xi ^{\mu }$
and $\sigma $. Therefore, as a rule, it turns out possible\textrm{\ }to find
in an explicit form some (or even all) nontrivial solutions of the DE
system, and thus to determine several (all) additional point Lie symmetry
groups inherent to the equation in question.

It should be remarked that various equations of form (\ref{BasicEqn1}) admit
only the point Lie groups generated by the vector fields (\ref{Kernel}) with 
$u(x)$ being any solution of the respective equation. For instance, it is
easy to check that all equations of the form (\ref{eq2.21}) such that $\chi
^{\alpha \beta }=\delta ^{\alpha \beta }$ and $\kappa (x)=p(x)$, where $p(x)$
is an arbitrary polynomial of $x^1$ and $x^2$, belong to this variety.
Without too much difficulties one can ascertain that the same holds true for
the equations of the form (\ref{Plate-Eqn}) with $D=const$, $N^{\alpha \beta
}=\delta ^{\alpha \beta }$ and $k(x)=p(x)$.

On the other hand, there are equations of the foregoing type which are
invariant under a larger group; an immediate example is the biharmonic
equation, $\Delta ^{2}w=0$, which admits the seven-parameter group generated
by the linear combinations of $X_{0}$ and the following six additional basic
vector fields (cf. Ovsiannikov, 1972): 
\begin{eqnarray*}
X_{1}=\frac{\partial }{\partial x^{1}},\quad X_{3}=x^{2}\frac{\partial }{%
\partial x^{1}}-x^{1}\frac{\partial }{\partial x^{2}},\quad X_{5}=2x^{1}x^{2}%
\frac{\partial }{\partial x^{1}}-\left[ \left( x^{1}\right) ^{2}-\left(
x^{2}\right) ^{2}\right] \frac{\partial }{\partial x^{2}}+2x^{2}w\frac{%
\partial }{\partial w}, \\
X_{2}=\frac{\partial }{\partial x^{2}},\quad X_{4}=x^{1}\frac{\partial }{%
\partial x^{1}}+x^{2}\frac{\partial }{\partial x^{2}},\quad X_{6}=\left[
\left( x^{1}\right) ^{2}-\left( x^{2}\right) ^{2}\right] \frac{\partial }{%
\partial x^{1}}+2x^{1}x^{2}\frac{\partial }{\partial x^{2}}+2x^{1}w\frac{%
\partial }{\partial w}.
\end{eqnarray*}
An important problem naturally arises in the light of the above note. It may
be placed in the category of the so-called group-classification problems
(see Ovsiannikov, 1982) and consist in determination of all those equations
of the type considered that admit a larger group together with this group
itself. Its most general statement assumes all functions $A^{\alpha \beta
\gamma \delta }(x),A^{\alpha \beta \gamma }(x),A^{\alpha \beta
}(x),A^{\alpha }(x),A(x),\xi ^{\alpha }(x)$ and $\sigma (x)$ involved in the
determining equations (\ref{eq3.14}) -- (\ref{eq3.18}) to be regarded as
unknown variables and to find all solutions of this system. Here we are not
going to study this rather complicated nonlinear problem in general.
However, in Sections 4 and 5, restricting our attention to the equations of
form (\ref{Plate-Eqn}), $D=const$, and (\ref{eq2.21}), respectively, we will
examine the corresponding group-classification problems.

Let us now specialize to the case of self-adjoint equations of form (\ref
{BasicEqn1}). Suppose that 
\begin{equation}
\mathcal{D}[w]=0,\quad \mathcal{D}=\mathcal{D}^{\ast },  \label{BasicEqn2}
\end{equation}
is such an equation. Then, of particular interest are its variational
symmetries -- the Lie groups generated by the so-called infinitesimal
divergence symmetries (see Definition 4.33 in Olver, 1993) of any
variational functional with (\ref{BasicEqn2}) as the associated
Euler-Lagrange equation. (Note that if two functionals lead to the same
Euler-Lagrange equation, then they have the same collection of infinitesimal
divergence symmetries.) This interest is motivated by the fact that, in
virtue of Noether's theorem, each variational symmetry of a given
self-adjoint equation corresponds to a conservation law admitted by the
smooth solutions of the equation. Thus, if a vector field $X$ of form (\ref
{eq3.2}) is found to generate a variational symmetry of equation (\ref
{BasicEqn2}), then Noether's theorem implies the existence of a conserved
current, which, in the present case, is a couple of differential functions $%
P^{\alpha }$ such that 
\begin{equation}
D_{\alpha }P^{\alpha }=Q\mathcal{D}[w],  \label{Cons-L-New}
\end{equation}
where $Q$ is the characteristic of $X$; by definition 
\begin{equation}
Q=\eta -w_{\mu }\xi ^{\mu }.  \label{Char-New}
\end{equation}
The total divergence of the conserved current $P^{\alpha }$ vanishes on the
smooth solutions of (\ref{BasicEqn2}) and so we have the conservation law 
\begin{equation}
D_{\alpha }P^{\alpha }=0,  \label{Cons-L-New-1}
\end{equation}
(\ref{Cons-L-New}) being its expression in characteristic form, and $Q$ --
its characteristics. Therefore, to derive the conservation laws of the
foregoing type, one can proceed by first determining the variational
symmetries of equation (\ref{BasicEqn2}), and than using their
characteristics (\ref{Char-New}) to find, from (\ref{Cons-L-New}), explicit
expressions for the corresponding conserved currents.

Having analyzed earlier the invariance properties of the whole class of
equations (\ref{BasicEqn1}), it is convenient to base the determination of
the variational symmetries of equation (\ref{BasicEqn2}) on the following
observation. A vector field $X$ of form (\ref{eq3.2}) generates a
variational symmetry of equation (\ref{BasicEqn2}) if and only if $X$ is an
infinitesimal symmetry of this equation, that is (\ref{eq3.3}) holds, and 
\begin{equation}
\mathrel{\mathop{X}\limits_{4}}\left( \mathcal{D}[w]\right) +\left( \frac{%
\partial \eta }{\partial w}+D_{\mu }\xi ^{\mu }\right) \mathcal{D}[w]=0.
\label{eq4.25}
\end{equation}
This is a consequence of Lemma 4.34 and Proposition 5.55 (Olver, 1993), see
also Lemma 7.46 (Olver, 1995). Subtracting (\ref{eq3.3}) from (\ref{eq4.25})
we can replace the latter with 
\[
\left( \frac{\partial \eta }{\partial w}+D_{\mu }\xi ^{\mu }+\lambda \right) 
\mathcal{D}[w]=0, 
\]
and as $\mathcal{D}[w]$ is not supposed to vanish identically we arrive at
the conclusion that 
\begin{equation}
\frac{\partial \eta }{\partial w}+D_{\mu }\xi ^{\mu }+\lambda =0,
\label{eq4.27-Gen}
\end{equation}
is a necessary and sufficient condition for an infinitesimal symmetry
admitted by a self-adjoint equation of form (\ref{BasicEqn1}) to be its
infinitesimal variational symmetry as well. It should be remarked that the
same holds true for any self-adjoint partial differential equation in one
dependent variable $w$ and $n$ independent variables $x=\left( x^{1},\ldots
,x^{n}\right) $; of course, in the general case the summation index $\mu $
will take the values $1,\ldots ,n$ in both formulae (\ref{eq3.2}) and (\ref
{eq4.27-Gen}). For a vector field of form (\ref{eq3.9}) the relation (\ref
{eq4.27-Gen}) simplifies, and reads 
\begin{equation}
\sigma +\xi _{,\mu }^{\mu }+\lambda =0.  \label{eq4.27}
\end{equation}

Thus to find the variational symmetries of an equation of form (\ref
{BasicEqn2}), it suffices to check which of its ''ordinary'' symmetries
satisfy the additional requirement (\ref{eq4.27-Gen}). For instance, the
result (i) implies that $X_{0}=w\partial /\partial w$ does not generate a
variational symmetry of any equation of form (\ref{BasicEqn2}), while a
vector field $X_{u}=u(x)\partial /\partial w$ generates a variational
symmetry of an equation of form (\ref{BasicEqn2}) whenever $u(x)$ is its
solution (this is a common property of all systems of linear homogeneous
partial differential equations, see Section 5.3 in Olver, 1993).

Suppose one has established that a vector field $X$ with characteristic $Q$
generates a variational symmetry of a given equation of form (\ref{BasicEqn2}%
), and now wishes to find the conserved current $P^{\alpha }$ of the
corresponding conservation law (\ref{Cons-L-New-1}). For this purpose one
can use formulae (5.150) and (5.151) given by Olver (1993) which express (in
an explicit form) a null Lagrangian as a divergence. Indeed, in this case
the right hand side of (\ref{Cons-L-New}) is a total divergence or, in other
words, a null Lagrangian. However, bearing in mind the recommendation of
Olver (1993) to use these formulae only as a last resort since ''the
homotopy formula (5.151) can rapidly become unmanageable'', in the present
paper we suggest another way for determination of the sought conserved
currents.

Our starting point is the so-called Noether identity (cf. Ibragimov, 1985): 
\begin{equation}
\mathrel{\mathop{X}\limits_{\infty }}(\mathcal{L)}+(D_{\alpha }\xi ^{\alpha
})\mathcal{L}=Q\mathsf{E}(\mathcal{L})+D_{\alpha }N^{\alpha }(\mathcal{L}),
\label{NoetId}
\end{equation}
which holds for any differential function $\mathcal{L}$ and vector field $X$
of the types considered here. In (\ref{NoetId}), $N^{\alpha }$ are the
differential operators given by the expressions 
\begin{eqnarray}
N^{\alpha } &=&\xi ^{\alpha }+Q\left\{ \frac{\partial }{\partial w_{\alpha }}%
+\sum\limits_{s\geq 1}(-1)^{s}D_{\nu _{1}}\cdots D_{\nu _{s}}\frac{\partial 
}{\partial w_{\alpha \nu _{1}\ldots \nu _{s}}}\right\}  \label{NoetOp} \\
&&\!\!\!\!\!\!\!\!\!\!+\dsum\limits_{r\geq 1}\left( D_{\mu _{1}}\cdots
D_{\mu _{r}}Q\right) \left\{ \frac{\partial }{\partial w_{\alpha \mu
_{1}\ldots \mu _{r}}}+\sum\limits_{s\geq 1}(-1)^{s}D_{\nu _{1}}\cdots D_{\nu
_{s}}\frac{\partial }{\partial w_{\alpha \mu _{1}\ldots \mu _{r}\nu
_{1}\ldots \nu _{s}}}\right\} ,  \nonumber
\end{eqnarray}
$Q=\eta -\xi ^{\alpha }w_{\alpha }$ being the characteristic of the vector
field $X$. Setting $\mathcal{L}=L$ in (\ref{NoetId}), and taking into
account (\ref{LD-1}) and (\ref{E-L-1}), after a little manipulation we
obtain the identity 
\begin{equation}
D_{\mu }N^{\mu }(-w\mathcal{D}[w])=-w\mathrel{\mathop{X}\limits_{4}}(%
\mathcal{D}[w])-\{\eta +(D_{\mu }\xi ^{\mu })w-2Q\}\mathcal{D}[w],
\label{DI}
\end{equation}
valid for any self-adjoint differential operator $\mathcal{D}$ of form (\ref
{DO}) and vector field of form (\ref{eq3.2}).

In particular, for $X_{v}=v(x)\partial /\partial w$, where $v(x)$ is an
arbitrary smooth function, we have 
\begin{equation}
\xi ^{\alpha }=0,\quad Q=\eta =v,  \label{DI-Lin1}
\end{equation}
and hence 
\begin{equation}
\mathrel{\mathop{X_v}\limits_{4}}\left( \mathcal{D}[w]\right) =\mathcal{D}%
[v],  \label{DI-Lin}
\end{equation}
since $\mathcal{D}$ is a linear differential operator. Substituting (\ref
{DI-Lin1}) and (\ref{DI-Lin}) into (\ref{DI}) we obtain 
\begin{equation}
D_{\mu }N^{\mu }(-w\mathcal{D}[w])=v\mathcal{D}[w]-w\mathcal{D}[v],
\label{RecId}
\end{equation}
which is nothing but the reciprocity relation associated with the equation $%
\mathcal{D}[w]=0$. Under the additional assumption $v=u(x)$, where $u(x)$ is
an arbitrary smooth solution of the latter equation, the reciprocity
relation (\ref{RecId}) becomes 
\begin{equation}
D_{\mu }N^{\mu }(-w\mathcal{D}[w])=u\mathcal{D}[w].  \label{ConsU}
\end{equation}
Taking into account (\ref{ConsU}) we can give now the following general
formula for the conserved currents $P^{\alpha }$ of the conservation laws
with characteristics $Q=u$ corresponding to the infinitesimal variational
symmetries $X_{u}=u\partial /\partial w$ of equation (\ref{BasicEqn2}): 
\[
P^{\alpha }=P_{(u)}^{\alpha }+G^{\alpha }, 
\]
where 
\begin{equation}
P_{(u)}^{\alpha }=N^{\alpha }(-wD[w]),  \label{NewCurrentU}
\end{equation}
and $G^{\alpha }$ is any null divergence. Of course, 
\[
D_{\mu }P_{(u)}^{\mu }=u\mathcal{D}[w], 
\]
and 
\begin{equation}
D_{\mu }P_{(u)}^{\mu }=0,  \label{ConsLowUa}
\end{equation}
on the smooth solutions of the equation (\ref{BasicEqn2}).

Next, let $X$ be an infinitesimal variational symmetry of equation (\ref
{BasicEqn2}) with characteristic $Q=w\sigma -w_{\mu }\xi ^{\mu }$. Then, on
account of (\ref{eq4.25}), (\ref{DI}) takes the form 
\[
D_{\mu }N^{\mu }\left( -\frac{1}{2}w\mathcal{D}[w]\right) =Q\mathcal{D}[w], 
\]
and hence we can write down the following explicit formula for the conserved
currents $P^{\alpha }$ of the conservation laws with characteristics $%
Q=w\sigma -w_{\mu }\xi ^{\mu }$ corresponding to the aforementioned
variational symmetries of equation (\ref{BasicEqn2}), namely 
\[
P^{\alpha }=B^{\alpha }+G^{\alpha }, 
\]
\begin{equation}
B^{\alpha }=N^{\alpha }(-\frac{1}{2}wD[w])+\frac{1}{2}D_{\mu }\left( w\xi
^{\alpha }A^{\mu \beta \gamma \delta }D_{\beta }D_{\gamma }D_{\delta }w-w\xi
^{\mu }A^{\alpha \beta \gamma \delta }D_{\beta }D_{\gamma }D_{\delta
}w\right) ,  \label{NewCurrent}
\end{equation}
where, as before, $G^{\alpha }$ is any null divergence. Of course, 
\[
D_{\mu }B^{\mu }=Q\mathcal{D}[w], 
\]
and on the smooth solutions of respective equation (\ref{BasicEqn2}) we have 
\[
D_{\mu }B^{\mu }=0, 
\]
Let us remark that the special null divergence 
\[
\frac{1}{2}D_{\mu }\left( w\xi ^{\alpha }A^{\mu \beta \gamma \delta
}D_{\beta }D_{\gamma }D_{\delta }w-w\xi ^{\mu }A^{\alpha \beta \gamma \delta
}D_{\beta }D_{\gamma }D_{\delta }w\right) , 
\]
is used in the expression (\ref{NewCurrent}) for the conserved current $%
B^{\alpha }$ to cut the fourth-order derivatives of the dependent variable $%
w $ away since in practice we are usually interested in conserved currents
which involve derivatives of order not higher than $k-1$, where $k$ is the
order of the equation considered. Making use of (\ref{NoetOp}) it is easy to
check that the right-hand side of (\ref{NewCurrent}) incorporates
derivatives of $w$ of order less than fourth. In the subsequent Sections
just (\ref{NewCurrentU}) and (\ref{NewCurrent}) will be referred to as the
expressions for the conserved currents of the conservation laws with
characteristics $Q=u$ and $Q=w\sigma -w_{\mu }\xi ^{\mu }$, respectively,
derived for equations of the form (\ref{BasicEqn2}).\pagebreak

To summarize, given an equation of form (\ref{BasicEqn2}), the crucial point
on the way of deriving conservation laws admitted by its smooth solutions is
to find vector fields of form (\ref{eq3.9}) generating ''ordinary'' point
Lie symmetries of the given equation. For that purpose, we should look for
solutions of the respective DE system (\ref{eq3.14}) -- (\ref{eq3.18}). Once
such vector fields are found, it is easy to check which of their linear
combinations satisfy the requirement (\ref{eq4.27-Gen}) and hence generate
variational symmetries of the equation considered. Now, using the
characteristics of these symmetries we first construct the operators $%
N^{\alpha }$ from formulae (\ref{NoetOp}) and then calculate from (\ref
{NewCurrent}) the conserved currents of the corresponding conservation laws.

\begin{center}
\bigskip

4. SYMMETRIES, CONSERVATION LAWS AND GROUP-INVARIANT SOLUTIONS OF PLATE
EQUATIONS\smallskip
\end{center}

In Section 2 (Example 1), we have quoted the self-adjoint equation (\ref
{Plate-Eqn}) describing the small bending of a plate resting on an elastic
foundation. Many problems concerning stability and vibration of isotropic
thin elastic plates are studied on the ground of this type of equations.
Here, we analyze the invariance properties of a generic equation of this
form under the assumption that the bending rigidity of the plates considered
is uniform, that is $D=const$. In this case (\ref{Plate-Eqn}) may be written
as follows 
\begin{equation}
A^{\alpha \beta \gamma \delta }w_{\alpha \beta \gamma \delta }+A^{\alpha
\beta }(x)w_{\alpha \beta }+A(x)w=0,  \label{Plate-Eqn1}
\end{equation}
with 
\begin{equation}
A^{\alpha \beta \gamma \delta }=\frac{1}{3}(\delta ^{\alpha \beta }\delta
^{\gamma \delta }+\delta ^{\alpha \gamma }\delta ^{\beta \delta }+\delta
^{\alpha \delta }\delta ^{\beta \gamma }),\quad A_{,\mu }^{\alpha \mu
}=0,\quad  \label{PE+}
\end{equation}
assuming that 
\[
A^{\alpha \beta }=\frac{1}{D}N^{\alpha \beta },\quad A=\frac{1}{D}k. 
\]

In view of the general results of Section 3, it is clear that $%
X_{u}=u(x)\partial /\partial w$ generates a variational symmetry of any
equation of form (\ref{Plate-Eqn1}) whenever $u(x)$ is its solution, while $%
X_{0}=w\partial /\partial w$ alone could never generate a variational
symmetry of an equation of form (\ref{Plate-Eqn1}), though it always is its
infinitesimal point Lie symmetry. Substituting (\ref{PE+}) into the
determining equations (\ref{eq3.14}) -- (\ref{eq3.18}) and taking into
account that $A^{\alpha \beta \gamma }=0$, $A^{\alpha }=0$, we obtain, after
a straightforward computation, that an equation of form (\ref{Plate-Eqn1})
is invariant under a point Lie group generated by a vector field $X$ of form
(\ref{eq3.9}), $X\neq cX_{0}$, if and only if 
\begin{equation}
\sigma =\frac{1}{2}\xi _{,\mu }^{\mu },  \label{DEs1}
\end{equation}
\begin{equation}
\delta ^{\alpha \mu }\xi _{,\mu }^{\beta }+\delta ^{\mu \beta }\xi _{,\mu
}^{\alpha }-\delta ^{\alpha \beta }\xi _{,\mu }^{\mu }=0,  \label{DEs3}
\end{equation}
\begin{equation}
\xi ^{\mu }A_{,\mu }^{\alpha \beta }-A^{\alpha \mu }\xi _{,\mu }^{\beta
}-A^{\mu \beta }\xi _{,\mu }^{\alpha }+2\xi _{,\mu }^{\mu }A^{\alpha \beta
}+2\delta ^{\alpha \tau }\delta ^{\beta \nu }\xi _{,\mu \tau \nu }^{\mu }=0,
\label{DEs4}
\end{equation}
\begin{equation}
A^{\alpha \nu }\xi _{,\mu \nu }^{\mu }-A^{\mu \nu }\xi _{,\mu \nu }^{\alpha
}=0,  \label{DEs5}
\end{equation}
\begin{equation}
2\xi ^{\mu }A_{,\mu }+4\xi _{,\mu }^{\mu }A+A^{\mu \nu }\xi _{,\tau \mu \nu
}^{\tau }=0.  \label{DEs6}
\end{equation}
At that, 
\begin{equation}
\lambda =-\frac{3}{2}\xi _{,\mu }^{\mu }.  \label{DEs2}
\end{equation}
Substituting expression (\ref{DEs1}) into (\ref{eq3.9}), and expressions (%
\ref{DEs1}) and (\ref{DEs2}) into condition (\ref{eq4.27}), we immediately
arrive at the conclusion that the generator of such a group is a vector
field of form 
\begin{equation}
X=\xi ^{\mu }\frac{\partial }{\partial x^{\mu }}+\frac{1}{2}\xi _{,\mu
}^{\mu }w\frac{\partial }{\partial w},  \label{VSGenerator}
\end{equation}
each such symmetry of (\ref{Plate-Eqn1}) being variational symmetry of the
latter equation as well, and hence there exist a conservation law with
characteristic $Q=(1/2)\xi _{,\mu }^{\mu }w-w_{\mu }\xi ^{\mu }$ and
conserved current $B^{\alpha }$ given by (\ref{NewCurrent}) admitted by the
smooth solutions of the equation considered. Thus to derive the conservation
laws, which correspond to the variational symmetries of an equation of form (%
\ref{Plate-Eqn1}) it suffices to know the results of the group
classification of the class of equations in question; of course, the same
holds true for the derivation of group-invariant solutions to (\ref
{Plate-Eqn1}). This group-classification problem is studied in Vassilev
(1988), (1991) and (1997). The classification results presented below are
obtained in these works.

It is shown that the scalar fields 
\begin{equation}
s_{\left( 1\right) }=A^{\mu \nu }\delta _{\mu \nu },\quad s_{\left( 2\right)
}=\left( 8A-\delta _{\alpha \mu }\delta _{\beta \nu }A^{\alpha \beta }A^{\mu
\nu }\right) ^{1/2},\quad s_{\left( 3\right) }=\left( -\delta ^{\mu \nu
}s_{\left( 1\right) ,\mu }s_{\left( 1\right) ,\nu }\right) ^{1/3},
\label{Invs}
\end{equation}
are of key importance for the group classification of the considered class
of equations. These scalar fields are called the invariants of equation (\ref
{Plate-Eqn1}) since here they play a role similar to the role that Laplace's
and Cotton's invariants play in the group classification of the second-order
linear partial differential equations (see Ovsiannikov, 1982 and Ibragimov,
1985). The following two properties of the scalar fields (\ref{Invs}) give
us both an additional reason to call them invariants of (\ref{Plate-Eqn1})
and explicit expressions for the invariants of groups admitted by (\ref
{Plate-Eqn1}). First, if an equation of form (\ref{Plate-Eqn1}) admits a
vector field of form (\ref{VSGenerator}), then 
\[
\xi _{,\mu }^{\mu }s_{(j)}+\xi ^{\mu }s_{(j),\mu }=0\quad (j=1,2,3), 
\]
and hence $U_{(j)}=w\sqrt{s_{(j)}}$ are invariants of the corresponding Lie
group whenever $s_{(j)}\not\equiv 0$. Second, if an equation of form (\ref
{Plate-Eqn1}) admits a vector field of form (\ref{VSGenerator}) and is such
that at least two of its invariants (\ref{Invs}), say $s_{(k)}$ and $%
s_{(l)}\;(k\neq l\leq 3)$ , are not identically equal to zero, then $%
s_{(k)}/s_{(l)}$ is an invariant of the corresponding symmetry group. Note,
that the invariants $s_{(k)}$ and $s_{(l)}\;$of such an equation of form (%
\ref{Plate-Eqn1}) provide two couples of functionally independent
invariants, namely $U_{(k)}=w\sqrt{s_{(k)}}$ and $s_{(k)}/s_{(l)}$ as well
as $U_{(l)}=w\sqrt{s_{(l)}}$ and $s_{(k)}/s_{(l)}$, of the admitted symmetry
group, both couples being readily applicable for constructing
group-invariant solutions to the respective equation. However, if even one
of the invariants (\ref{Invs}) of an equation of form (\ref{Plate-Eqn1}) is
not identically equal to zero, then this equation admits at most a
3-parameter group with generators of form (\ref{VSGenerator}). On the other
hand, if all invariants (\ref{Invs}) of an equation of form (\ref{Plate-Eqn1}%
) are identically equal to zero, then this equation admits a 6-parameter
group with generators of form (\ref{VSGenerator}). Below, the latter case is
set out in detail.

Let $\omega (z)\not\equiv const$ be an analytic function of the complex
variable $z=x^{1}+ix^{2}$, and let $E_{\omega }$ be the equation of the form
(\ref{Plate-Eqn1}) with coefficients 
\begin{equation}
A^{11}=-A^{22}=4\func{Re}\left\{ \phi \right\} ,\quad A^{12}=A^{21}=-4\func{%
Im}\left\{ \phi \right\} ,\quad A=4\phi \bar{\phi},  \label{Coeff}
\end{equation}
where $\phi $ is the Schwarzian derivative of the function $\omega $, that
is 
\begin{equation}
\phi =\left( \frac{\omega ^{\prime \prime }}{\omega ^{\prime }}\right)
^{\prime }-\frac{1}{2}\left( \frac{\omega ^{\prime \prime }}{\omega ^{\prime
}}\right) ^{2},  \label{Schwarz}
\end{equation}
$\bar{\phi}$ is the complex conjugated of $\phi $, and the prime is used to
denote differentiation with respect to the variable $z$. Substituting (\ref
{Coeff}) into (\ref{Invs}) one can see that all invariants of $E_{\omega }$
are identically equal to zero. Then, taking into account the DE system (\ref
{DEs3})--(\ref{DEs6}), (\ref{Coeff}) and (\ref{Schwarz}), one can verify by
direct computing that $E_{\omega }$ admits the 6-parameter group generated
by the vector fields 
\[
Z_{(j)}=\xi _{(j)}^{\mu }\frac{\partial }{\partial \,x^{\mu }}+\frac{1}{2}%
\xi _{(j),\mu }^{\mu }w\frac{\partial }{\partial w}\quad (j=1,\ldots ,6), 
\]
the functions $\xi _{(j)}^{\mu }$ being given by the expressions 
\[
\begin{tabular}{ll}
$\xi _{(1)}^{1}=\func{Re}\left\{ \omega _{1}\right\} ,$ & $\xi _{(1)}^{2}=%
\func{Im}\left\{ \omega _{1}\right\} ,\smallskip $ \\ 
$\xi _{(2)}^{1}=\func{Re}\left\{ i\omega _{1}\right\} ,$ & $\xi _{(2)}^{2}=%
\func{Im}\left\{ i\omega _{1}\right\} ,\smallskip $ \\ 
$\xi _{(3)}^{1}=\func{Re}\left\{ \omega _{2}\right\} ,$ & $\xi _{(3)}^{2}=%
\func{Im}\left\{ \omega _{2}\right\} ,\smallskip $ \\ 
$\xi _{(4)}^{1}=\func{Re}\left\{ i\omega _{2}\right\} ,$ & $\xi _{(4)}^{2}=%
\func{Im}\left\{ i\omega _{2}\right\} ,\smallskip $ \\ 
$\xi _{(5)}^{1}=\func{Re}\left\{ \omega _{3}\right\} ,$ & $\xi _{(5)}^{2}=%
\func{Im}\left\{ \omega _{3}\right\} ,\smallskip $ \\ 
$\xi _{(6)}^{1}=\func{Re}\left\{ i\omega _{3}\right\} ,$ & $\xi _{(6)}^{2}=%
\func{Im}\left\{ i\omega _{3}\right\} .$%
\end{tabular}
\]
where 
\begin{equation}
\omega _{1}=\frac{1}{\omega ^{\prime }},\quad \omega _{2}=\frac{\omega }{%
\omega ^{\prime }},\quad \omega _{3}=\frac{\omega ^{2}}{\omega ^{\prime }}.
\label{func}
\end{equation}
It should be remarked that each equation of form (\ref{Plate-Eqn1}) which
admits a 6-parameter group with generators of form (\ref{VSGenerator}) is of
type $E_{\omega }$, meaning that it can be generated in the above manner
using a suitable analytic function $\omega $. The coefficients of each
equation of this type are of the form $A^{\alpha \beta }=\delta ^{\alpha \mu
}\delta ^{\beta \nu }\mathit{\varphi }_{,\mu \nu }$, $A=(1/8)\delta ^{\alpha
\mu }\delta ^{\beta \nu }\mathit{\varphi }_{,\alpha \beta }\mathit{\varphi }%
_{,\mu \nu }$, where $\mathit{\varphi }$ is a harmonic function, that is $%
\delta ^{\alpha \beta }\mathit{\varphi }_{,\alpha \beta }=0$, and vice
versa. It is noteworthy that each equation with variable coefficients of
type $E_{\omega }$ can be mapped to an equation with constant coefficients
belonging to the same family. It is easy to verify by direct computing that
the equation $E_{\omega }$ corresponding to an analytic function $\omega $
whose Schwarzian derivative is not constant transforms to a constant
coefficients one under the following change of the independent and dependent
variables: 
\begin{equation}
y^{\alpha }=f^{\alpha }\left( x^{1},x^{2}\right) ,\;W=wU\left(
x^{1},x^{2}\right) ,  \label{NewVar}
\end{equation}
\[
f^{1}\left( x^{1},x^{2}\right) =\func{Re}\left\{ \int f^{-1}dz\right\}
,\;f^{2}\left( x^{1},x^{2}\right) =\func{Im}\left\{ \int f^{-1}dz\right\}
,\;U\left( x^{1},x^{2}\right) =\left( f\stackrel{-}{f}\right) ^{-1/2}, 
\]
where $f$ is any linear combination of the functions (\ref{func}) such that $%
f\not\equiv 0$, i.e. 
\begin{equation}
f=k_{1}\omega _{1}+k_{2}\omega _{2}+k_{3}\omega _{3},  \label{NewVarF}
\end{equation}
where $k_{1}$, $k_{2}$, and $k_{3}$ are complex constants such that $%
k_{1}^{2}+k_{2}^{2}+k_{3}^{2}\neq 0$.

Consider, as a simple example, the equation $E_{\omega }$ corresponding to $%
\omega =z^{\sqrt{\varkappa /2}}$, where $\varkappa $ is a positive real
constant. In this case (\ref{Schwarz}) gives $\phi =4\left( 2-\varkappa
\right) z^{2}$ and hence, according to (\ref{Coeff}), the coefficients of $%
E_{\omega }$ read 
\[
A^{11}=-A^{22}=\left( 2-\varkappa \right) \frac{\left( x^{1}\right)
^{2}-\left( x^{2}\right) ^{2}}{\left[ \left( x^{1}\right) ^{2}+\left(
x^{2}\right) ^{2}\right] ^{2}},\quad A^{12}=A^{21}=\left( 2-\varkappa
\right) \frac{2x^{1}x^{2}}{\left[ \left( x^{1}\right) ^{2}+\left(
x^{2}\right) ^{2}\right] ^{2}}, 
\]
\begin{equation}
A=\left( 2-\varkappa \right) ^{2}\frac{1}{4\left[ \left( x^{1}\right)
^{2}+\left( x^{2}\right) ^{2}\right] ^{2}}.  \label{ExCoeff}
\end{equation}
Using the function $f=z$ obtained from (\ref{NewVarF}) for $\omega =z^{\sqrt{%
\varkappa /2}}$, $k_{1}=k_{3}=0$, $k_{2}=1+\sqrt{\varkappa /2}$, we
introduce, according to (\ref{NewVar}), the new independent and dependent
variables 
\begin{equation}
y^{1}=\frac{1}{2}\ln \left[ \left( x^{1}\right) ^{2}+\left( x^{2}\right) ^{2}%
\right] ,\quad y^{2}=\arctan \left( \frac{x^{2}}{x^{1}}\right) ,\quad W=w%
\left[ \left( x^{1}\right) ^{2}+\left( x^{2}\right) ^{2}\right] ^{-1/2}.
\label{NewVarEx}
\end{equation}
Note that the inverse transformations are given by the expressions 
\begin{equation}
x^{1}=e^{y^{1}}\cos y^{2},\quad x^{2}=e^{y^{1}}\sin y^{2},\quad w=W\left[
\left( x^{1}\right) ^{2}+\left( x^{2}\right) ^{2}\right] ^{1/2}.
\label{NewVarExI}
\end{equation}
Under the change of the variables according to (\ref{NewVarEx}), the
considered equation $E_{\omega }$ transforms to the following one, 
\begin{equation}
\delta ^{\alpha \beta }\delta ^{\mu \nu }\frac{\partial ^{4}W}{\partial
y^{\alpha }\partial y^{\beta }\partial y^{\mu }\partial y^{\nu }}-\varkappa 
\frac{\partial ^{2}W}{\partial y^{1}\partial y^{1}}+\varkappa \frac{\partial
^{2}W}{\partial y^{2}\partial y^{2}}+\frac{1}{4}\varkappa ^{2}W=0,
\label{TEq}
\end{equation}
which belongs to the same class (since it corresponds to the analytic
function $\omega =e^{z\sqrt{\varkappa /2}}$) but whose coefficients are
constant.

Equation (\ref{TEq}) admits the 6-parameter group of variational symmetries
generated by the basic vector fields 
\begin{eqnarray*}
V_{\alpha } &=&\frac{\partial }{\partial y^{\alpha }}, \\
V_{3} &=&e^{\theta y^{1}}\cos \left( \theta y^{2}\right) \frac{\partial }{%
\partial y^{1}}+e^{\theta y^{1}}\sin \left( \theta y^{2}\right) \frac{%
\partial }{\partial y^{2}}+\theta e^{\theta y^{1}}w\cos \left( \theta
y^{2}\right) \frac{\partial }{\partial w}, \\
V_{4} &=&-e^{\theta y^{1}}\cos \left( \theta y^{2}\right) \frac{\partial }{%
\partial y^{1}}+e^{\theta y^{1}}\cos \left( \theta y^{2}\right) \frac{%
\partial }{\partial y^{2}}-\theta e^{\theta y^{1}}w\sin \left( \theta
y^{2}\right) \frac{\partial }{\partial w}, \\
V_{5} &=&e^{-\theta y^{1}}\cos \left( \theta y^{2}\right) \frac{\partial }{%
\partial y^{1}}-e^{-\theta y^{1}}\sin \left( \theta y^{2}\right) \frac{%
\partial }{\partial y^{2}}-\theta e^{-\theta y^{1}}w\cos \left( \theta
y^{2}\right) \frac{\partial }{\partial w}, \\
V_{6} &=&e^{-\theta y^{1}}\cos \left( \theta y^{2}\right) \frac{\partial }{%
\partial y^{1}}+e^{-\theta y^{1}}\sin \left( \theta y^{2}\right) \frac{%
\partial }{\partial y^{2}}-\theta e^{\theta y^{1}}w\cos \left( \theta
y^{2}\right) \frac{\partial }{\partial w},
\end{eqnarray*}
where $\theta =\sqrt{\varkappa /2}$. These vector fields give rise to six
linearly independent conservation laws for equation (\ref{TEq}). The
characteristics of these conservation laws are 
\[
Q_{\left( j\right) }=\frac{1}{2}W\frac{\partial }{\partial y^{\mu }}%
V_{j}\left( y^{\mu }\right) -W_{\mu }V_{j}\left( y^{\mu }\right) \quad
\left( j=1,\ldots ,6\right) . 
\]
Here, $V_{j}$ are regarded as operators acting on the functions $\zeta
:R^{2}\rightarrow R$. The corresponding conserved currents can be easily
calculated from (\ref{NewCurrent}).

Finally, let us remark that each one-parameter group generated by a linear
combination of the basic vector fields $V_{j}$ can be used for constructing
group-invariant solutions of equation (\ref{TEq}). Consider, for instance,
the group $H\left( V_{3}+V_{5}\right) $ generated by the vector field $%
V_{3}+V_{5}$. The functions $s=\sin \left( \theta y^{2}\right) /\cosh \left(
\theta y^{1}\right) $ and $u=W/\cosh \left( \theta y^{1}\right) $ constitute
a complete set of invariants for this group and hence, following the well
known algorithm (Ovsiannikov, 1982; Olver, 1993), we seek the $H\left(
V_{3}+V_{5}\right) $--invariant solutions of equation (\ref{TEq}) in the
form 
\begin{equation}
W=u\left( s\right) \cosh \left( \theta y^{1}\right) ,\quad s=\frac{\sin
\left( \theta y^{2}\right) }{\cosh \left( \theta y^{1}\right) }.
\label{InvSol1}
\end{equation}
Substituting (\ref{InvSol1}) into (\ref{TEq}), we get the reduced equation 
\[
\left( s^{2}-1\right) ^{2}\frac{d^{4}u}{ds^{4}}+8s\left( s^{2}-1\right) 
\frac{d^{3}u}{ds^{3}}+4\left( 3s^{2}-1\right) \frac{d^{2}u}{ds^{2}}=0. 
\]
The general solution to this ordinary differential equation is 
\[
u\left( s\right) =C_{1}+C_{2}\ln \left( \frac{s+1}{s-1}\right)
+C_{3}s+C_{4}s\ln \left( \frac{s+1}{s-1}\right) , 
\]
where $C_{1}$, $C_{2}$, $C_{3}$ and $C_{4}$ are real constants. Hence the $%
H\left( V_{3}+V_{5}\right) $--invariant solutions of equation (\ref{TEq})
are given by the expression 
\[
W\left( y^{1},y^{2}\right) =C_{1}\cosh \left( \theta y^{1}\right)
+C_{2}\cosh \left( \theta y^{1}\right) \ln \left[ \frac{\sin \left( \theta
y^{2}\right) +\cosh \left( \theta y^{1}\right) }{\sin \left( \theta
y^{2}\right) -\cosh \left( \theta y^{1}\right) }\right] 
\]
\[
\qquad \qquad \quad +\;C_{3}\sin \left( \theta y^{2}\right) +C_{4}\frac{\sin
\left( \theta y^{2}\right) }{\cosh \left( \theta y^{1}\right) }\ln \left[ 
\frac{\sin \left( \theta y^{2}\right) +\cosh \left( \theta y^{1}\right) }{%
\sin \left( \theta y^{2}\right) -\cosh \left( \theta y^{1}\right) }\right] . 
\]
Using the inverse transformations (\ref{NewVarExI}) one can convert the
above solutions of equation (\ref{TEq}) into solutions of the equation $%
E_{\omega }$, $\omega =z^{\sqrt{\varkappa /2}}$, with variable coefficients (%
\ref{ExCoeff}).\pagebreak

\begin{center}
4. SYMMETRIES, CONSERVATION LAWS AND GROUP-INVARIANT SOLUTIONS OF ROD
EQUATIONS\smallskip
\end{center}

In Section 2, combining and generalizing Examples 3 and 4 we have introduced
the class of self-adjoint partial differential equations 
\begin{equation}
A^{1111}w_{1111}+A^{\alpha \beta }w_{\alpha \beta }+Aw=0,  \label{eq5.1}
\end{equation}
with coefficients 
\begin{equation}
A^{1111}=\gamma ,\quad A^{\alpha \beta }=\chi ^{\alpha \beta },\quad
A=\kappa (x),  \label{eq1coef}
\end{equation}
where $\gamma =const\neq 0$, $\chi ^{\alpha \beta }$ are arbitrary constants
(but $\left( \chi ^{12}\right) ^{2}+\left( \chi ^{22}\right) ^{2}\neq 0$,
otherwise (\ref{eq5.1}) degenerates and becomes an ordinary differential
equation), and $\kappa (x)$ is an arbitrary function. Equations of this
special type are used by many authors to study applied engineering problems
concerning dynamics and stability of both elastic beams resting on elastic
foundations (see e.g. Smith and Herrmann, 1972) and pipes conveying fluid
(see e.g. Gregory and Paidoussis, 1966). In the present Section we first
examine the point Lie symmetries of (\ref{eq5.1}) and solve the
corresponding group-classification problem. Then we derive conservation laws
and group-invariant solutions of various rod equations of form (\ref{eq5.1}).

Consider the group-classification problem. In view of the results (i) and
(ii) of Section 3, it is clear that each equation of form (\ref{eq5.1}) is
invariant under the point Lie groups generated by the vector fields $%
X_0=w\partial /\partial w$ and $X_u=u\left( x\right) \partial /\partial w$
where $u\left( x\right) $ is any smooth solution of the respective equation
and the objective is to find those equations of the type considered which
admit vector fields $X$ of form (\ref{eq3.9}), $X\neq cX_0$, $c=const\neq 0$.

Substituting (\ref{eq1coef}) into (\ref{eq3.14}) -- (\ref{eq3.18}), and
taking into account that $\,A^{\alpha },\,A^{\alpha \beta \gamma }$ and all $%
A^{\alpha \beta \gamma \delta }$ except $A^{1111}$ are equal to zero, we
obtain after a little manipulation the following system of determining
equations for the functions $\xi ^{\mu }\left( x\right) $ and $\sigma \left(
x\right) $ associated with the sought vector fields $X$ of form (\ref{eq3.9}%
): 
\begin{equation}
\xi _{,1}^{2}=0,\quad 2\sigma _{,1}-3\xi _{,11}^{1}=0,  \label{a}
\end{equation}
\begin{equation}
5\gamma \xi _{,111}^{1}+2\chi ^{11}\xi _{,1}^{1}-2\chi ^{12}\xi _{,2}^{1}=0,
\label{eq5.4.2}
\end{equation}
\begin{equation}
\chi ^{22}(2\xi _{,1}^{1}-\xi _{,2}^{2})=0,\quad \chi ^{12}(3\xi
_{,1}^{1}-\xi _{,2}^{2})-\chi ^{22}\xi _{,2}^{1}=0,  \label{b}
\end{equation}
\begin{equation}
2\chi ^{12}\sigma _{,2}-\chi ^{22}\xi _{,22}^{1}=0,\quad 2\chi ^{12}\sigma
_{,1}+\chi ^{22}(2\sigma _{,2}-\xi _{,22}^{2})=0,  \label{eq5.4.4}
\end{equation}
\begin{equation}
\sigma _{,1111}+\chi ^{\alpha \beta }\sigma _{,\alpha \beta }+\xi ^{\mu
}\kappa _{,\mu }+4\xi _{,1}^{1}\kappa =0,  \label{eq5.4.5}
\end{equation}
the auxiliary function $\lambda $ being expressed by 
\begin{equation}
\lambda =\sigma -4\xi _{,1}^{1}.  \label{lambda}
\end{equation}
We look for the equations of form (\ref{eq5.1}) whose coefficients $\gamma $%
, $\chi ^{\alpha \beta }$ and $\kappa \left( x\right) $ are such that system
(\ref{a}) -- (\ref{eq5.4.5}) possesses solutions different from the trivial
one $\xi ^{\mu }=0$, $\sigma =const\neq 0$.

Observing the system of determining equations (\ref{a}) -- (\ref{eq5.4.5})
we see that for the purposes of the group-classification it is convenient to
divide the equations of form (\ref{eq5.1}) into three subclasses depending
on whether the coefficients $\chi ^{\alpha \beta }$ of the given equation
are such that: (A) $\chi ^{22}\neq 0,\det (\chi ^{\alpha \beta })\neq 0$;
(B) $\chi ^{22}\neq 0,\det (\chi ^{\alpha \beta })=0$ or (C) $\chi
^{22}=0,\det (\chi ^{\alpha \beta })\neq 0$. This covers all possibilities
except for $\chi ^{22}=0,\det (\chi ^{\alpha \beta })=0$ when (\ref{eq5.1})
becomes an ordinary differential equation that falls outside our interest in
the present paper.

For convenience we introduce the notation 
\[
Y_{\alpha }=\frac{\partial }{\partial x^{\alpha }},\;Y_{3}=\left( x^{1}+%
\frac{\chi ^{12}}{\chi ^{22}}x^{2}\right) \frac{\partial }{\partial x^{1}}%
+2x^{2}\frac{\partial }{\partial x^{2}},\;Y_{4}=\left( x^{1}+\frac{\chi ^{11}%
}{\chi ^{12}}x^{2}\right) \frac{\partial }{\partial x^{1}}+3x^{2}\frac{%
\partial }{\partial x^{2}}. 
\]

Let $\chi ^{22}\neq 0$. Then, the first equation (\ref{b}) is equivalent to 
\begin{equation}
2\xi _{,1}^{1}-\xi _{,2}^{2}=0.  \label{eq5.5}
\end{equation}
Differentiating (\ref{eq5.5}) with respect to $x^{1}$ and taking into
account the first equation (\ref{a}) we obtain $\xi _{,11}^{1}=0$. Hence, (%
\ref{eq5.4.2}) and the second equation (\ref{b}) reduce to 
\begin{equation}
\chi ^{11}\xi _{,1}^{1}-\chi ^{12}\xi _{,2}^{1}=0,\quad \chi ^{12}\xi
_{,1}^{1}-\chi ^{22}\xi _{,2}^{1}=0.  \label{eq5.5.1}
\end{equation}

Consider the subclass (A): $\chi ^{22}\neq 0,$ $\det (\chi ^{\alpha \beta
})\neq 0$. Then, the first equation (\ref{a}) together with (\ref{eq5.5})
and (\ref{eq5.5.1}) lead to $\xi _{,1}^{1}=\xi _{,2}^{1}=\xi _{,1}^{2}=\xi
_{,2}^{2}=0$, i.e., $\xi ^{\alpha }=c^{\alpha }=const$. Consequently, the
second equations of (\ref{a}) and (\ref{eq5.4.4}) imply $\sigma =c=const$.
At this, the first equation (\ref{eq5.4.4}) is satisfied and (\ref{eq5.4.5})
becomes 
\begin{equation}
c^{\alpha }\kappa _{,\alpha }=0.  \label{c}
\end{equation}
All this means that when $\chi ^{22}\neq 0$ and $\det (\chi ^{\alpha \beta
})\neq 0$ the DE system has only the trivial solution unless 
\begin{equation}
\kappa (x)=f(\beta ^{2}x^{1}-\beta ^{1}x^{2}),  \label{eq5.5.2}
\end{equation}
for a certain smooth function $f\not\equiv const$ and certain constants $%
\beta ^{\alpha }$ such that $\left( \beta ^{1}\right) ^{2}+\left( \beta
^{1}\right) ^{2}\neq 0$. In this latter case, the DE system has the
nontrivial solution 
\begin{equation}
\xi ^{1}=\beta ^{1},\quad \xi ^{2}=\beta ^{2},\quad \sigma =0,  \label{eq5.6}
\end{equation}
and so the differential equations of that kind admit additionally the
one-parameter symmetry group associated with the vector field $\beta
^{1}Y_{1}+\beta ^{2}Y_{2}$. In the case $\kappa =const$, (\ref{c}) is
satisfied for any couple of constants $c^{\alpha }$, and hence such
equations admit the 2-parameter symmetry group with generators $Y_{1}$ and $%
Y_{2}$. This completes the analysis of subclass (A).

Consider now the subclass (B): $\chi ^{22}\neq 0,\det (\chi ^{\alpha \beta
})=0$. Differentiating the second equation (\ref{eq5.5.1}) successively with
respect to $x^{1}$ and $x^{2}$ we obtain 
\[
\chi ^{12}\xi _{,11}^{1}-\chi ^{22}\xi _{,12}^{1}=0,\quad \chi ^{12}\xi
_{,12}^{1}-\chi ^{22}\xi _{,22}^{1}=0, 
\]
which, on account of $\xi _{,11}^{1}=0$, leads to $\xi _{,12}^{1}=\xi
_{,22}^{1}=0$. This result, together with (\ref{eq5.5}) imply $\xi
_{,22}^{2}=0$, and the nontrivial solution of (\ref{a}) -- (\ref{eq5.4.4})
is now obvious: 
\begin{equation}
\xi ^{1}=c^{1}+c^{3}\left( x^{1}+\frac{\chi ^{12}}{\chi ^{22}}x^{2}\right)
,\quad \xi ^{2}=c^{2}+2c^{3}x^{2},\quad \sigma =0,  \label{eq5.7}
\end{equation}
with $c^{1},\ c^{2}$ and $c^{3}$ -- arbitrary constants. Equation (\ref
{eq5.4.5}) reduces to 
\begin{equation}
\xi ^{\alpha }\kappa _{,\alpha }+4\xi _{,1}^{1}\kappa =0.  \label{clasify1}
\end{equation}
For an arbitrary $\kappa (x)$ it leads to $\xi ^{\alpha }=0$; it is easily
verified that for $\kappa (x)=x^{1}+x^{2}x^{2}$, $\xi ^{\alpha }=0$ is the
only solution of (\ref{clasify1}).

If $\kappa (x)$ is a function of form (\ref{eq5.5.2}), then (\ref{clasify1})
implies (\ref{eq5.6}) as a nontrivial solution to the system (\ref{a}) -- (%
\ref{eq5.4.5}). Therefore, a generic equation of this kind admits only the
one-parameter group generated by $\beta ^{1}Y_{1}+\beta ^{2}Y_{2}$, unless $%
\kappa (x)$ is of one of the following two special forms. The first one is 
\begin{equation}
\kappa (x)=\kappa _{0}\left( \beta +x^{2}\right) ^{-2},\quad \kappa
_{0}=const\neq 0,\quad \beta =const,  \label{eq5.8.1}
\end{equation}
when the nontrivial solution of (\ref{a}) -- (\ref{eq5.4.5}) is (\ref{eq5.7}%
) with $c^{2}=2\beta $, $c^{3}=1$, $c^{1}$ -- arbitrary, and hence the
equations of subclass (B) with $\kappa (x)$ of form (\ref{eq5.8.1}) admit
the 2-parameter symmetry group generated by the vector fields $Y_{1}$ and $%
2\beta Y_{2}+Y_{3}$. The second special form of the function $\kappa (x)$ is 
\begin{equation}
\kappa (x)=\kappa _{0}\left( \beta +x^{1}-\frac{\chi ^{12}}{\chi ^{22}}%
x^{2}\right) ^{-4},\quad \kappa _{0}=const\neq 0,\quad \beta =const,
\label{eq5.8.3}
\end{equation}
when the respective differential equations admit the 2-parameter group
spanned over the vector fields $\beta Y_{1}+Y_{3}$ and $\left( \chi
^{12}/\chi ^{22}\right) Y_{1}+Y_{2}$.

Another extension of the symmetry group is possible if there exist two
constants $\beta ^{1}$ and $\beta ^{2}$, as well as a smooth function $%
f\not\equiv 0$ such that 
\begin{equation}
\kappa (x)=\left( \beta ^{2}+x^{2}\right) ^{-2}f(y),\quad y=\left( \beta
^{2}+x^{2}\right) ^{-1/2}\left( \beta ^{1}+x^{1}-\frac{\chi ^{12}}{\chi ^{22}%
}x^{2}\right) .  \label{eq5.8.2}
\end{equation}
If $\kappa (x)$ is of form (\ref{eq5.8.2}), then the nontrivial solution of
the determining equations (\ref{a}) -- (\ref{eq5.4.5}) is (\ref{eq5.7}) with 
$c^{1}=\beta ^{1}+2\left( \chi ^{12}/\chi ^{22}\right) \beta ^{2}$, $%
c^{2}=2\beta ^{2}$, $c^{3}=1$ and the differential equations of that sort
admit the one-parameter group, generated by 
\[
\left( \beta ^{1}+2\frac{\chi ^{12}}{\chi ^{22}}\beta ^{2}\right)
Y_{1}+2\beta ^{2}Y_{2}+Y_{3} 
\]
only, except for the cases $f(y)=\kappa _{0}y^{-4}$ ($\kappa _{0}=const$),
when $\kappa (x)$ takes the form (\ref{eq5.8.3}), and $f(y)=const\neq 0$
when $\kappa (x)$ becomes (\ref{eq5.8.1}).

The differential equations of form (\ref{eq5.1}) with $\chi ^{22}\neq 0$, $%
\det (\chi ^{\alpha \beta })=0$ admit the 2-parameter group generated by $%
Y_{1}$ and $Y_{2}$ when $\kappa (x)=const\neq 0$ or the 3-parameter group
with generators $Y_{1}$, $Y_{2}$ and $Y_{3}$ when $\kappa (x)=0$.

Finally, consider the subclass (C): $\chi ^{22}=0,\det (\chi ^{\alpha \beta
})\neq 0$. Substituting $\chi ^{22}=0$ in the determining equations (\ref{a}%
) -- (\ref{eq5.4.4}), the latter simplify to 
\[
\xi _{,1}^{2}=0,\quad \xi _{,11}^{1}=0,\quad \chi ^{11}\xi _{,1}^{1}-\chi
^{12}\xi _{,2}^{1}=0,\quad 3\xi _{,1}^{1}-\xi _{,2}^{2}=0,\quad \sigma
_{,1}=0,\quad \sigma _{,2}=0, 
\]
and their nontrivial solution is easily obtained: 
\begin{equation}
\xi ^{1}=c^{1}+c^{3}\left( x^{1}+\frac{\chi ^{11}}{\chi ^{12}}x^{2}\right)
,\quad \xi ^{2}=c^{2}+3c^{3}x^{2},\quad \sigma =0,  \label{eq5.9}
\end{equation}
where $c^{i}$ are arbitrary constants (note that if $\chi ^{22}=0,$ then $%
\det (\chi ^{\alpha \beta })\neq 0$ assumes $\chi ^{12}\neq 0$). Equation (%
\ref{eq5.4.5}) takes the form (\ref{clasify1}) and for an arbitrary $\kappa
(x)$ it leads to ${\xi ^{\alpha }}=0$. If $\kappa (x)$ is of the form (\ref
{eq5.5.2}), such equations admit the one-parameter group generated by $\beta
^{1}Y_{1}+\beta ^{2}Y_{2}$ only, except for two special forms of $\kappa (x)$%
, namely 
\begin{equation}
\kappa (x)=\kappa _{0}\left( \beta +x^{2}\right) ^{-4/3},\quad
\label{eq5.9.1}
\end{equation}
and 
\begin{equation}
\kappa (x)=\kappa _{0}\left( \beta +2x^{1}-\frac{\chi ^{11}}{\chi ^{12}}%
x^{2}\right) ^{-4}.  \label{eq5.10.1}
\end{equation}
where $\kappa _{0}$ and $\beta $ are constants. In the case (\ref{eq5.9.1}),
the nontrivial solution of the determining equation (\ref{a}) -- (\ref
{eq5.4.5}) is (\ref{eq5.9}) with $c^{2}=3\beta $, $c^{3}=1$, $c^{1}$ --
arbitrary, and the differential equation considered admits the 2-parameter
group spanned over the vector fields $Y_{1}$ and $3\beta Y_{2}+Y_{4}$. In
the case (\ref{eq5.10.1}) the group admitted is also a 2-parameter one, but
generated by $\beta Y_{1}+Y_{4}$ and $\left( \chi ^{11}/\chi ^{12}\right)
Y_{1}+2Y_{2}$.

Another extension of the symmetry group is possible if 
\begin{equation}
\kappa (x)=\left( \beta ^{2}+x^{2}\right) ^{-4/3}f(y),\quad y=\left( \beta
^{2}+x^{2}\right) ^{-1/3}\left( \beta ^{1}+2x^{1}-\frac{\chi ^{11}}{\chi
^{12}}x^{2}\right) ,  \label{eq5.10}
\end{equation}
where $\beta ^{\alpha }$ are constants and $f\not\equiv \;0$ is a smooth
function. In this case, the nontrivial solution of (\ref{a}) -- (\ref
{eq5.4.5}) is (\ref{eq5.9}) with $c^{1}=\beta ^{1}+3(\chi ^{11}/\chi
^{12})\beta ^{2}$, $c^{2}=6\beta ^{2}$, $c^{3}=1$ and we conclude that the
equations of subclass (C) with $\kappa (x)$ of form (\ref{eq5.10}) admit
only the one-parameter group generated by $[\beta ^{1}+3(\chi ^{11}/\chi
^{12})\beta ^{2}]Y_{1}+6\beta ^{2}Y_{2}+Y_{4}$, except for $f(y)=\kappa
_{0}y^{-4}$ ($\kappa _{0}=const$), when (\ref{eq5.10}) coincides with (\ref
{eq5.10.1}) or $f(y)=const\neq 0$ when $\kappa (x)$ has the form (\ref
{eq5.9.1}).

Evidently, the differential equations of form (\ref{eq5.1}) with $\chi
^{22}=0$, $\det (\chi ^{\alpha \beta })\neq 0$ admit: the 2-parameter group
with generators $Y_{1}$ and $Y_{2}$ when $\kappa (x)=const\neq 0$, and the
3-parameter group associated with $Y_{1}$, $Y_{2}$ and $Y_{4}$ -- when $%
\kappa (x)=0$.

The results of the above group-classification analysis are summarized in
Table 1, where the equations invariant under larger groups are given through
their coefficients together with the generators of the associated symmetry
groups.

\begin{center}
Table 1. Equations of form (\ref{eq5.1}) invariant under larger symmetry
groups.\smallskip

\begin{tabular}{ccc}
\hline\hline
\# & Coefficients & Generators \\ \hline\hline
\multicolumn{1}{l}{} & \multicolumn{1}{l}{} & \multicolumn{1}{l}{} \\[-9pt] 
\multicolumn{1}{l}{1} & \multicolumn{1}{l}{$\;\,\kappa (x)=f(\beta
^{2}x^{1}-\beta ^{1}x^{2})$} & \multicolumn{1}{l}{$\;\,\beta ^{1}Y_{1}+\beta
^{2}Y_{2}$} \\[2pt] \hline
\multicolumn{1}{l}{} & \multicolumn{1}{l}{} & \multicolumn{1}{l}{} \\[-9pt] 
\multicolumn{1}{l}{2} & \multicolumn{1}{l}{$
\begin{array}{l}
\chi ^{22}\neq 0,\;\det (\chi ^{\alpha \beta })=0,\;\kappa (x)=\left( \beta
^{2}+x^{2}\right) ^{-2}f(y), \\ 
y=\left( \beta ^{2}+x^{2}\right) ^{-1/2}[\beta ^{1}+x^{1}-\left( \chi
^{12}/\chi ^{22}\right) x^{2}]
\end{array}
$} & \multicolumn{1}{l}{$
\begin{array}{l}
\lbrack \beta ^{1}+2(\chi ^{12}/\chi ^{22})\beta ^{2}]Y_{1} \\ 
+2\beta ^{2}Y_{2}+Y_{3}
\end{array}
$} \\[13pt] \hline
\multicolumn{1}{l}{} & \multicolumn{1}{l}{} & \multicolumn{1}{l}{} \\[-9pt] 
\multicolumn{1}{l}{3} & \multicolumn{1}{l}{$
\begin{array}{l}
\chi ^{22}=0,\;\det (\chi ^{\alpha \beta })\neq 0,\;\kappa (x)=\left( \beta
^{2}+x^{2}\right) ^{-4/3}f(y), \\ 
y=\left( \beta ^{2}+x^{2}\right) ^{-1/3}[\beta ^{1}+2x^{1}-\left( \chi
^{11}/\chi ^{12}\right) x^{2}]
\end{array}
$} & \multicolumn{1}{l}{$
\begin{array}{l}
\lbrack \beta ^{1}+3(\chi ^{11}/\chi ^{12})\beta ^{2}]Y_{1} \\ 
+6\beta ^{2}Y_{2}+2Y_{4}
\end{array}
$} \\[13pt] \hline
\multicolumn{1}{l}{} & \multicolumn{1}{l}{} & \multicolumn{1}{l}{} \\[-9pt] 
\multicolumn{1}{l}{4} & \multicolumn{1}{l}{$\;\,\chi ^{22}\neq 0,\;\det
(\chi ^{\alpha \beta })=0,\;\kappa (x)=\kappa _{0}\left( \beta +x^{2}\right)
^{-2},$} & \multicolumn{1}{l}{$\;\,Y_{1}$, $2\beta Y_{2}+Y_{3}$} \\%
[2pt] \hline
\multicolumn{1}{l}{} & \multicolumn{1}{l}{} & \multicolumn{1}{l}{} \\[-9pt] 
\multicolumn{1}{l}{5} & \multicolumn{1}{l}{$\;\,\chi ^{22}=0,\;\det (\chi
^{\alpha \beta })\neq 0,\;\kappa (x)=\kappa _{0}\left( \beta +x^{2}\right)
^{-4/3}$} & \multicolumn{1}{l}{$\;\,Y_{1}$, $3\beta Y_{2}+Y_{4}$} \\%
[2pt] \hline
\multicolumn{1}{l}{} & \multicolumn{1}{l}{} & \multicolumn{1}{l}{} \\[-9pt] 
\multicolumn{1}{l}{6} & \multicolumn{1}{l}{$
\begin{array}{l}
\chi ^{22}\neq 0,\;\det (\chi ^{\alpha \beta })=0, \\ 
\kappa (x)=\kappa _{0}\left( \beta +x^{1}-\left( \chi ^{12}/\chi
^{22}\right) x^{2}\right) ^{-4}
\end{array}
$} & \multicolumn{1}{l}{$
\begin{array}{l}
\beta Y_{1}+Y_{3}, \\ 
(\chi ^{12}/\chi ^{22})Y_{1}+Y_{2}
\end{array}
$} \\[11pt] \hline
\multicolumn{1}{l}{} & \multicolumn{1}{l}{} & \multicolumn{1}{l}{} \\[-9pt] 
\multicolumn{1}{l}{7} & \multicolumn{1}{l}{$
\begin{array}{l}
\chi ^{22}=0,\;\det (\chi ^{\alpha \beta })\neq 0, \\ 
\kappa (x)=\kappa _{0}\left( \beta +2x^{1}-\left( \chi ^{11}/\chi
^{12}\right) x^{2}\right) ^{-4}
\end{array}
$} & \multicolumn{1}{l}{$
\begin{array}{l}
\beta Y_{1}+2Y_{4}, \\ 
(\chi ^{11}/\chi ^{12})Y_{1}+2Y_{2}
\end{array}
$} \\[11pt] \hline
\multicolumn{1}{l}{} & \multicolumn{1}{l}{} & \multicolumn{1}{l}{} \\[-9pt] 
\multicolumn{1}{l}{8} & \multicolumn{1}{l}{$\;\,\chi ^{22}\det (\chi
^{\alpha \beta })\neq 0,\;\kappa (x)=const$} & \multicolumn{1}{l}{$\;\,Y_{1}$%
, $Y_{2}$} \\[2pt] \hline
\multicolumn{1}{l}{} & \multicolumn{1}{l}{} & \multicolumn{1}{l}{} \\[-9pt] 
\multicolumn{1}{l}{9} & \multicolumn{1}{l}{$\;\,\chi ^{22}\det (\chi
^{\alpha \beta })=0,\;\kappa (x)=const\neq 0$} & \multicolumn{1}{l}{$%
\;\,Y_{1}$, $Y_{2}$} \\[2pt] \hline
\multicolumn{1}{l}{} & \multicolumn{1}{l}{} & \multicolumn{1}{l}{} \\[-9pt] 
\multicolumn{1}{l}{10} & \multicolumn{1}{l}{$\;\,\chi ^{22}\neq 0,\;\det
(\chi ^{\alpha \beta })=0,\;\kappa (x)=0$} & \multicolumn{1}{l}{$\;\,Y_{1}$, 
$Y_{2}$, $Y_{3}$} \\[2pt] \hline
\multicolumn{1}{l}{} & \multicolumn{1}{l}{} & \multicolumn{1}{l}{} \\[-9pt] 
\multicolumn{1}{l}{11} & \multicolumn{1}{l}{$\;\,\chi ^{22}=0,\;\det (\chi
^{\alpha \beta })\neq 0,\;\kappa (x)=0$} & \multicolumn{1}{l}{$\;\,Y_{1}$, $%
Y_{2}$, $Y_{4}$} \\[2pt] \hline\hline
\end{tabular}
\smallskip
\end{center}

Having completely solved the group-classification problem, our next step is
to identify the variational symmetries of those equations of form (\ref
{eq5.1}) which have been found to admit larger symmetry groups. For this
purpose, we are to apply the condition (\ref{eq4.27-Gen}) to the linear
combinations of (\ref{Kernel}) and the vector fields presented in Table 1,
the respective functions $\lambda $ being given by (\ref{lambda}). Thus, we
found that all vector fields quoted under numbers 1, 3, 5, 7, 8, 9 and 11
generate variational symmetries as well. In case \# 2 the variational
symmetries are associated with $[\beta ^{1}+2(\chi ^{12}/\chi ^{22})\beta
^{2}]Y_{1}+2\beta ^{2}Y_{2}+Y_{3}+(1/2)X_{0}$. Similarly, in case \# 4 the
variational symmetries are generated by $Y_{1}$ and $2\beta
Y_{2}+Y_{3}+(1/2)X_{0}$, in case \# 6 -- by $(\chi ^{12}/\chi
^{22})Y_{1}+Y_{2}$ and $\beta Y_{1}+Y_{3}+(1/2)X_{0}$, and in case \# 10 --
by $Y_{1}$, $Y_{2}$ and $Y_{3}+(1/2)X_{0}$.

Once the variational symmetries of the equations (\ref{eq5.1}) are
identified, we can derive the corresponding conservation laws. The conserved
currents of the conservation laws for the equations given in Table 1 are
computed using (\ref{NewCurrent}), the above notes concerning the
corresponding variational symmetries being taken into account. The obtained
conservation laws are listed in Table 2 (in the same order as in Table 1) in
terms of the differential functions: 
\begin{eqnarray*}
B_{(1)}^{1} &=&-\frac{1}{2}[\gamma (2w_{1}w_{111}-w_{11}^{2})+\chi
^{11}w_{1}^{2}-\chi ^{22}w_{2}^{2}+\kappa w^{2}]-\frac{1}{2}(\chi ^{2\mu
}ww_{\mu }),_{2}, \\
B_{(1)}^{2} &=&-\chi ^{2\mu }w_{1}w_{\mu }+\frac{1}{2}(\chi ^{2\mu }ww_{\mu
}),_{1}, \\
B_{(2)}^{1} &=&-\chi ^{1\mu }w_{2}w_{\mu }+\gamma
(w_{11}w_{12}-w_{2}w_{111})-\frac{1}{2}(\gamma w_{1}w_{11}-\chi ^{1\mu
}ww_{\mu }),_{2}, \\
B_{(2)}^{2} &=&-\frac{1}{2}\left( \gamma w_{11}^{2}+\chi ^{22}w_{2}^{2}-\chi
^{11}w_{1}^{2}+\kappa w^{2}\right) +\frac{1}{2}(\gamma w_{1}w_{11}-\chi
^{1\mu }ww_{\mu }),_{1}, \\
B_{(3)}^{\alpha } &=&\left( x^{1}+\frac{\chi ^{12}}{\chi ^{22}}x^{2}\right)
B_{(1)}^{\alpha }+2x^{2}B_{(2)}^{\alpha }+\chi ^{\alpha \mu }ww_{\mu }+\frac{%
1}{2}\gamma \delta ^{1\alpha }(ww_{111}-w_{1}w_{11}), \\
B_{(4)}^{\alpha } &=&\left( x^{1}+\frac{\chi ^{11}}{\chi ^{12}}x^{2}\right)
B_{(1)}^{\alpha }+3x^{2}B_{(2)}^{\alpha } \\
&&+\frac{1}{2}[\chi ^{\alpha \mu }ww_{\mu }+\delta ^{1\alpha }(\chi
^{11}ww_{1}+2\chi ^{12}ww_{2}-\gamma w_{1}w_{11})].
\end{eqnarray*}

\begin{center}
Table 2. Conservation laws\smallskip\ for equations of form (\ref{eq5.1})

\begin{tabular}{cc}
\hline\hline
\# & Conservation laws \\ \hline\hline
\multicolumn{1}{l}{} & \multicolumn{1}{l}{} \\[-9pt] 
\multicolumn{1}{l}{1} & \multicolumn{1}{l}{$D_{\alpha }[\beta
^{1}B_{(1)}^{\alpha }+\beta ^{2}B_{(2)}^{\alpha }]=0$} \\[6pt] \hline
\multicolumn{1}{l}{} & \multicolumn{1}{l}{} \\[-9pt] 
\multicolumn{1}{l}{2} & \multicolumn{1}{l}{$D_{\alpha }[\left( \beta
^{1}+2(\chi ^{12}/\chi ^{22})\beta ^{2}\right) B_{(1)}^{\alpha }+2\beta
^{2}B_{(2)}^{\alpha }+B_{(3)}^{\alpha }]=0$} \\[7pt] \hline
\multicolumn{1}{l}{} & \multicolumn{1}{l}{} \\[-9pt] 
\multicolumn{1}{l}{3} & \multicolumn{1}{l}{$D_{\alpha }[\left( \beta
^{1}+3(\chi ^{11}/\chi ^{12})\beta ^{2}\right) B_{(1)}^{\alpha }+6\beta
^{2}B_{(2)}^{\alpha }+2B_{(4)}^{\alpha }]=0$} \\[7pt] \hline
\multicolumn{1}{l}{} & \multicolumn{1}{l}{} \\[-9pt] 
\multicolumn{1}{l}{4} & \multicolumn{1}{l}{$D_{\alpha }B_{(1)}^{\alpha }=0$%
,\quad $D_{\alpha }[2\beta B_{(2)}^{\alpha }+B_{(3)}^{\alpha }]=0$} \\%
[6pt] \hline
\multicolumn{1}{l}{} & \multicolumn{1}{l}{} \\[-9pt] 
\multicolumn{1}{l}{5} & \multicolumn{1}{l}{$D_{\alpha }B_{(1)}^{\alpha }=0$%
,\quad $D_{\alpha }[3\beta B_{(2)}^{\alpha }+B_{(4)}^{\alpha }]=0$} \\%
[6pt] \hline
\multicolumn{1}{l}{} & \multicolumn{1}{l}{} \\[-9pt] 
\multicolumn{1}{l}{6} & \multicolumn{1}{l}{$D_{\alpha }[\beta
B_{(1)}^{\alpha }+B_{(3)}^{\alpha }]=0$,\quad $D_{\alpha }[(\chi ^{12}/\chi
^{22})B_{(1)}^{\alpha }+B_{(2)}^{\alpha }]=0$} \\[7pt] \hline
\multicolumn{1}{l}{} & \multicolumn{1}{l}{} \\[-9pt] 
\multicolumn{1}{l}{7} & \multicolumn{1}{l}{$D_{\alpha }[\beta
B_{(1)}^{\alpha }+2B_{(4)}^{\alpha }]=0$,\quad $D_{\alpha }[(\chi ^{11}/\chi
^{12})B_{(1)}^{\alpha }+2B_{(2)}^{\alpha }]=0$} \\[7pt] \hline
\multicolumn{1}{l}{} & \multicolumn{1}{l}{} \\[-9pt] 
\multicolumn{1}{l}{8} & \multicolumn{1}{l}{$D_{\alpha }B_{(1)}^{\alpha }=0$%
,\quad $D_{\alpha }B_{(2)}^{\alpha }=0$} \\[4pt] \hline
\multicolumn{1}{l}{} & \multicolumn{1}{l}{} \\[-9pt] 
\multicolumn{1}{l}{9} & \multicolumn{1}{l}{$D_{\alpha }B_{(1)}^{\alpha }=0$%
,\quad $D_{\alpha }B_{(2)}^{\alpha }=0$} \\[4pt] \hline
\multicolumn{1}{l}{} & \multicolumn{1}{l}{} \\[-9pt] 
\multicolumn{1}{l}{10} & \multicolumn{1}{l}{$D_{\alpha }B_{(1)}^{\alpha }=0$%
,\quad $D_{\alpha }B_{(2)}^{\alpha }=0$,\quad $D_{\alpha }B_{(3)}^{\alpha
}=0 $} \\[4pt] \hline
\multicolumn{1}{l}{} & \multicolumn{1}{l}{} \\[-9pt] 
\multicolumn{1}{l}{11} & \multicolumn{1}{l}{$D_{\alpha }B_{(1)}^{\alpha }=0$%
,\quad $D_{\alpha }B_{(2)}^{\alpha }=0$,\quad $D_{\alpha }B_{(4)}^{\alpha
}=0 $} \\[4pt] \hline\hline
\end{tabular}
\end{center}

According to the general results of Section 3, each equation (\ref{eq5.1})
admits conservation laws with characteristics $Q=u(x)$, where $u(x)$ is any
smooth solution of the equation considered. These conservation laws are of
the form (\ref{ConsLowUa}), that is 
\[
D_{\alpha }P_{(u)}^{\alpha }=0, 
\]
the corresponding conserved currents $P_{(u)}^{\alpha }$ being given by the
expression (\ref{NewCurrentU}). Here, on account of (\ref{eq1coef}), (\ref
{NewCurrentU}) simplifies and reads 
\begin{equation}
P_{(u)}^{\alpha }=\chi ^{\alpha \mu }(uw_{\mu }-u,_{\mu }w)+\delta ^{1\alpha
}\gamma (uw_{111}+u,_{11}w_{1}-u,_{111}w-u,_{1}w_{11}).  \label{eq5.12}
\end{equation}

Let us now specialize to the differential equation 
\begin{equation}
EJw_{1111}+mw_{22}=0,  \label{eq5.13}
\end{equation}
governing the dynamics of a classic homogeneous Bernoulli-Euler beam. Here $%
EJ$ is the bending rigidity of the beam and $m$ is the mass of the beam per
unit length. According to the above analysis, (\ref{eq5.13}) admits the
following six linearly independent infinitesimal variational symmetries: 
\begin{equation}
Y_{1},\quad Y_{2},\quad Y_{3}+\frac{1}{2}X_{0},\quad Y_{5}=\frac{\partial }{%
\partial w},\quad Y_{6}=x^{1}\frac{\partial }{\partial w},\quad Y_{7}=x^{2}%
\frac{\partial }{\partial w},  \label{eq5.14}
\end{equation}
where $Y_{5}$, $Y_{6}$ and $Y_{7}$ are vector fields of the type $%
X_{u}=u\left( x\right) \partial /\partial w$ corresponding to the solutions $%
u=1$, $u=x^{1}$ and $u=x^{2}$ of (\ref{eq5.13}), respectively. Here, the
independent variables $x^{1}$ and $x^{2}$ are the spatial variable along the
rod axis and the time, respectively, so that the conservation laws admitted
by the smooth solutions of equation (\ref{eq5.13}) may be written in the
more familiar form 
\[
\frac{\partial \mathit{\Psi }}{\partial x^{2}}+\frac{\partial P}{\partial
x^{1}}=0, 
\]
where $\mathit{\Psi }$ and $P$ denote the density and flux of the
conservation law, respectively. The densities and fluxes of the conservation
laws for (\ref{eq5.13}) associated with the vector fields (\ref{eq5.14})
together with their physical interpretation are presented in Table 3.

\begin{center}
Table 3. Conservation laws for Bernoulli-Euler beams\smallskip

\begin{tabular}{cc}
\hline\hline
Generators & Conservation laws \\ \hline\hline
$
\begin{array}{c}
\text{space translations} \\ 
Y_{1}
\end{array}
$ & \multicolumn{1}{l}{$
\begin{array}{l}
\text{\textbf{wave momentum}} \\ 
\Psi _{(1)}=mw_{1}w_{2} \\ 
P_{(1)}=(1/2)[EJ(2w_{1}w_{111}-w_{11}^{2})-mw_{2}^{2}]
\end{array}
$} \\[15pt] \hline
$
\begin{array}{c}
\text{time translations} \\ 
Y_{2}
\end{array}
$ & \multicolumn{1}{l}{$
\begin{array}{l}
\text{\textbf{energy}} \\ 
\Psi _{(2)}=(1/2)\left( EJw_{11}^{2}+mw_{2}^{2}\right) \\ 
P_{(2)}=EJ(w_{11}w_{12}-w_{2}w_{111})
\end{array}
$} \\[15pt] \hline
& \multicolumn{1}{l}{} \\[-9pt] 
$
\begin{array}{c}
\text{scaling} \\ 
Y_{3}+(1/2)X_{0}
\end{array}
$ & \multicolumn{1}{l}{$
\begin{array}{l}
\Psi _{(3)}=x^{1}\Psi _{(1)}+2x^{2}\Psi _{(2)}+mww_{2} \\ 
\multicolumn{1}{c}{
P_{(3)}=x^{1}P_{(1)}+2x^{2}P_{(2)}+(1/2)EJ(ww_{111}-w_{1}w_{11})}
\end{array}
$} \\[10pt] \hline
$Y_{5}$ & \multicolumn{1}{l}{$
\begin{array}{l}
\text{\textbf{linear momentum}} \\ 
\Psi _{(5)}=mw_{2},\quad P_{(5)}=EJw_{111}
\end{array}
$} \\[10pt] \hline
$Y_{6}$ & \multicolumn{1}{l}{$
\begin{array}{l}
\text{\textbf{similar to Eshelby energy-momentum tensor}} \\ 
\Psi _{(6)}=x^{1}mw_{2},\quad P_{(6)}=EJ(x^{1}w_{111}-w_{11})
\end{array}
$} \\[10pt] \hline
$
\begin{array}{c}
\text{Galilean boost} \\ 
Y_{7}
\end{array}
$ & \multicolumn{1}{l}{$
\begin{array}{l}
\text{\textbf{center-of-mass theorem}} \\ 
\Psi _{(7)}=m(x^{2}w_{2}-w),\quad P_{(7)}=EJx^{2}w_{111}
\end{array}
$} \\[10pt] \hline\hline
\end{tabular}
\end{center}

Conservation laws in the dynamics of rods are considered in many papers (see
e.g. Antman, 1984; Kienzler, 1986; Chien et al., 1993; Maddocks and
Dichmann, 1994; Tabarrok et al., 1994; Djondjorov, 1995). However, the
particular form of the differential equations examined in the present study
allows comparison with the results reported by Chien et al. (1993) and by
Maddocks and Dichmann (1994) only.

Chien et al. (1993) derive conservation laws for the statics and dynamics of
rods employing a technique called by the authors Neutral Action (NA) method.
The conservation laws for rod equations established in this Section could be
compared to their ones only for the differential equation (\ref{eq5.13})
which coincides with the equation 
\[
Bw_{1111}+2B_{,1}w_{111}+B_{,11}w_{11}+Hw_{22}=0, 
\]
considered in Chien et al. (1993) when $B=EJ$ and $H=m$. The comparison
shows that the conserved currents of the conservation laws for (\ref{eq5.13}%
) with characteristics other than $Q=u(x)$ obtained by Chien et al. (1993)
coincide with ours presented in Table 3. As for the conserved currents of
the conservation laws for (\ref{eq5.13}) with characteristics $Q=u(x)$,
where $u(x)$ is any solution of (\ref{eq5.13}), our general formula (\ref
{eq5.12}) implies 
\[
P_{(u)}^{\alpha }=\delta ^{2\alpha }m(uw_{2}-u,_{2}w)+\delta ^{1\alpha
}EJ(uw_{111}+u,_{11}w_{1}-u,_{111}w-u,_{1}w_{11}). 
\]
Only a part of these conservation laws are identified and presented in
(Chien et al., 1993), namely those associated with the solutions to (\ref
{eq5.13}) of the form 
\[
u(x)=C_{1}(x^{1})^{3}+C_{2}(x^{1})^{2}+C_{3}x^{1}+C_{4}+C_{5}x^{2},\quad
C_{i}=const\quad \left( i=1,\ldots ,5\right) , 
\]
while, in fact, equation (\ref{eq5.13}) has an infinite-dimensional space of
solutions.

Five conservation laws in the dynamics of rods are reported in (Maddocks and
Dichmann, 1994) within a general nonlinear direct theory. The restricted
version of this theory describing small planar bending of an uniform
inextensible unshearable isotropic elastic rod with a linear constitutive
law, the rotatory inertia of the rod cross section being neglected, is
exactly the classic Bernoulli-Euler theory for homogeneous beams whose
governing equation is (\ref{eq5.13}). Rewriting the conservation laws in
(Maddocks and Dichmann, 1994) taking into account the aforementioned
restrictions we observe that: (\textit{1}) the conservation law for the
total angular momentum (formula 2.14 in Maddocks and Dichmann, 1994)
degenerates to the well known basic relation of Bernoulli-Euler theory $%
Q=\partial M/\partial x^{1}$ (here $Q$ and $M$ denote shear force and
bending moment, respectively, see Washizu, 1982); (\textit{2}) the density
and flux of the conservation law associated with material isotropy (formula
4.5 in Maddocks and Dichmann, 1994) vanish identically; (\textit{3}) the
conservation law corresponding to material homogeneity (formula 3.2 in
Maddocks and Dichmann, 1994) reduces to conservation of the wave momentum
(see Table 3); (4) the expressions for the densities and fluxes of energy
(formula 2.19 in Maddocks and Dichmann, 1994) and linear momentum (formula
2.12 in Maddocks and Dichmann, 1994) conservation laws coincide with the
respective ones presented in Table 3. The set of conservation laws with
characteristics $Q=u(x)$, where $u(x)$ is any solution of (\ref{eq5.13}), as
well as the conservation law associated with the variational scaling
symmetry $Y_{3}+(1/2)X_{0}$ (see Table 3) have no analogues in (Maddocks and
Dichmann, 1994).

Three interesting kinds of group-invariant solutions to certain equations of
the class (\ref{eq5.1}) are identified below. First of them corresponds to
vector fields $cY_{1}\mp Y_{2},$ where $c=const$. These group-invariant
solutions are travelling waves 
\[
w=U(s),\quad s=x^{1}\pm cx^{2}, 
\]
admissible only for equations (\ref{eq5.1}) with $\kappa (x^{1},x^{2})=f(s)$%
. The reduced equations determining such group-invariant solutions are 
\[
\gamma \frac{d^{4}U}{ds^{4}}+(\chi ^{11}\pm 2\chi ^{12}c+\chi ^{22}c^{2})%
\frac{d^{2}U}{ds^{2}}+f(s)U=0. 
\]
The second one corresponds to the vector field $Y_{3}$ and is of the form 
\[
w=U(s),\quad s=x^{1}(x^{2})^{-1/2}-\frac{\chi ^{12}}{\chi ^{22}}%
(x^{2})^{1/2}. 
\]
The vector field $Y_{3}$ is admitted only if $\kappa
(x^{1},x^{2})=(x^{2})^{-2}f(s)$ (see cases \# 2, 4, 6 and 10 in Table 1).
The reduced equations for these invariant solutions are 
\[
4\gamma \frac{d^{4}U}{ds^{4}}+\chi ^{22}s^{2}\frac{d^{2}U}{ds^{2}}+3\chi
^{22}s\frac{dU}{ds}+4f(s)U=0. 
\]
The third kind of group-invariant solutions corresponds to the vector field $%
Y_{4}$: 
\[
w=U(s),\quad s=2x^{1}(x^{2})^{-1/3}-\frac{\chi ^{11}}{\chi ^{12}}%
(x^{2})^{2/3}. 
\]
The vector field $Y_{4}$ is admitted only if $\kappa
(x^{1},x^{2})=(x^{2})^{-4/3}f(s)$ (see cases \# 3, 5, 7 and 11 in Table 1).
The reduced equations for the invariant solution under consideration are 
\[
48\gamma \frac{d^{4}U}{ds^{4}}-4\chi ^{12}s\frac{d^{2}U}{ds^{2}}-4\chi ^{12}%
\frac{dU}{ds}+3f(s)U=0. 
\]
Obviously, the latter two kinds of group-invariant solutions could be
reduced to self-similar solutions if $\chi ^{12}=0$ or $\chi ^{11}=0$,
respectively.

\begin{center}
\medskip

\medskip 6. CONCLUDING REMARKS
\end{center}

In this paper, Lie transformation group methods have been applied to the
class of partial differential equations (\ref{BasicEqn}). This class is of
interest for structural mechanics since the governing equations of various
classical plate and rod theories belong to it; the examples given in Section
2 illustrate this fact. In the context of structural mechanics, the results
of the group analysis of equations (\ref{BasicEqn}) give a number of
attractive possibilities. Here, the established point Lie symmetries of (\ref
{BasicEqn}) are used to construct group-invariant solution to the governing
equations of several plate and rod models, to derive conservation laws
revealing important features of such models and to find transformations
simplifying the differential structure of equations associated with
particular plate problems.

First of all, the well known computational procedure for finding the most
general point Lie symmetry group has been applied to the foregoing class of
equations. As a result, the system of equations (\ref{eq3.14}) -- (\ref
{eq3.18}) is derived determining the equations of the type considered that
admit a larger group together with the generators of this group; naturally,
all equations of this class being linear and homogeneous admit the point Lie
groups generated by the vector fields (\ref{Kernel}). The system (\ref
{eq3.14}) -- (\ref{eq3.18}) allows the associated group-classification
problem to be stated and examined.

In Section 4, this problem is solved for the plate equations (\ref
{Plate-Eqn1}) in terms of their invariants $s_{\left( 1\right) }$, $%
s_{\left( 2\right) }$ and $s_{\left( 3\right) }$ defined by (\ref{Invs}).
The equations of form (\ref{Plate-Eqn1}) with $s_{\left( 1\right) }\equiv
s_{\left( 2\right) }\equiv s_{\left( 3\right) }\equiv 0$ are found to admit
the largest symmetry groups. It is noteworthy that each such equation with
variable coefficients can be transformed, using a suitable change of
variables, to an equation with constant coefficients belonging to the same
class. An example of such a transformation is given at the end of Section 4
where, in addition, a class of group-invariant solutions to the equation
considered is presented.

The group-classification problem for the rod equations (\ref{eq5.1}) is
entirely solved in Section~5. All equations of that kind admitting point Lie
symmetry groups, in addition to the ones generated by (\ref{Kernel}), are
determined and presented in Table 1 together with the generators of the
respective groups. The largest symmetry groups are admitted by the equations
of form (\ref{eq5.1}) whose coefficients are such that $\chi ^{22}\det (\chi
^{\alpha \beta })=0$, $\kappa (x)\equiv 0$. The most interesting
group-invariant solutions for equations (\ref{eq5.1}) are identified and the
corresponding reduced equations are presented at the end of Section 5.

Once the ''ordinary'' point Lie symmetries of an equation of form (\ref
{BasicEqn}) are determined, one can easily find, using the general criterion
(\ref{eq4.27-Gen}), which of them are variational symmetries of this
equation. Then, (\ref{NewCurrentU}) and (\ref{NewCurrent}) provide explicit
expressions for the conserved currents of the conservation laws associated
through Noether's theorem with the established variational symmetries. These
expressions will involve derivatives of the dependent variable of lowest
possible order, which is important in view of their application in
structural mechanics. The reciprocity relation valid for each equation of
form (\ref{BasicEqn}) is given explicitly by formula (\ref{RecId}).

In Section 4, it is shown, using the consequence (\ref{eq4.27}) of the
general criterion (\ref{eq4.27-Gen}), that each point Lie symmetry of a\
plate equation of form (\ref{Plate-Eqn1}) generated by a vector field of
form (\ref{VSGenerator}) is variational symmetry of this equation. Therefor,
each such symmetry gives rise to a conservation law with characteristic $%
Q=(1/2)\xi _{,\mu }^{\mu }w-w_{\mu }\xi ^{\mu }$ and conserved current given
by (\ref{NewCurrent}) admitted by the smooth solutions of the respective
equation.

The conservation laws for the\ rod equations listed in Table 1 are given in
Table 2. Inspecting these results one can see that the equations for
unsupported rods and rods on Winkler foundations admit two independent
conservation laws associated with the wave momentum ($D_{\alpha
}B_{(1)}^{\alpha }=0$) and energy ($D_{\alpha }B_{(2)}^{\alpha }=0$).
Equations (\ref{eq2.12}) and (\ref{eq2.16}) governing the stability of
unsupported axially compressed beams and fluid conveying pipes belong to
this class. Rod equations with $\kappa (x)=0$ and $\det (\chi ^{\alpha \beta
})=0$ admit a supplementary conservation law $D_{\alpha }B_{(3)}^{\alpha }=0$
associated with the infinitesimal scaling symmetry $Y_{3}$. Such an equation
is (\ref{eq5.13}) governing the vibration of the classic Bernoulli-Euler
beam. Table 3 contains several physically important conservation laws for
this equation. A comparison between the conservation laws derived here for
equation (\ref{eq5.13}) and the relevant results in (Chien et al., 1993) and
(Maddocks and Dichmann, 1994) is presented in Section 5.

\bigskip

\noindent \textit{Acknowledgements}---This research was supported by
Contract MM 517/1995 with the NSF, Bulgaria.

\begin{center}
\medskip

\medskip REFERENCES
\end{center}

\begin{description}
\item  Antman, S.S., 1984. The theory of rods. In: Mechanics of solids, Vol.
II. Springer-Verlag, Berlin, pp. 641-703.\vspace{-6pt}

\item  Bluman, G.W., Kumei, S., 1989. Symmetries and Differential Equations.
Springer-Verlag, New York.\vspace{-6pt}

\item  Chien, N., Honein, T., Herrmann, G., 1993. Conservation laws for
nonhomogeneous Bernoulli-Euler beams. International Journal of Solids and
Structures 30 (23), 3321-3335.\vspace{-6pt}

\item  Djondjorov, P., 1995. Invariant properties of Timoshenko beams
equations. International Journal of Engineering Science 33 (14), 2103-2114.%
\vspace{-6pt}

\item  Gregory, R.W., Paidoussis, M.P., 1966. Unstable oscillation of
tubular cantilevers conveying fluid. I: Theory. Proceedings of Royal Society
London A--293, 512-527.\vspace{-6pt}

\item  Ibragimov, N.H., 1985. Transformation Groups Applied to Mathematical
Physics. Reidel, Boston.\vspace{-6pt}

\item  Kienzler, R., 1986. On existence and completeness of conservation
laws associated with elementary beam theory. International Journal of Solids
and Structures 22 (7), 789-796.\vspace{-6pt}

\item  Maddocks, J., Dichmann, D., 1994. Conservation laws in the dynamics
of rods. Journal of Elasticity 34, 83-96.\vspace{-6pt}

\item  Olver, P.J., 1993. Applications of Lie Groups to Differential
Equations, Second Edition, Graduate Texts in Mathematics, Vol. 107.
Springer-Verlag, New York.\vspace{-6pt}

\item  Olver, P.J., 1995. Equivalence, Invariance and Symmetry. Cambridge
University Press, Cambridge.\vspace{-6pt}

\item  Ovsiannikov, L.V., 1972. Group properties of the equations of
mechanics. In: Mechanics of Continuous Media and Relevant Problems of
Analysis. Nauka, Moscow, pp. 381-393.\vspace{-6pt}

\item  Ovsiannikov, L.V., 1982. Group Analysis of Differential Equations.
Academic Press, New York.\vspace{-6pt}

\item  Smith, T.E., Herrmann, G., 1972. Stability of a beam on an elastic
foundation subjected to follower forces. Journal of Sound and Vibration 39,
628-629.\vspace{-6pt}

\item  Tabarrok, B., Tezer, C., Stylianou, M., 1994. A note on conservation
principles in classical mechanics. Acta Mechanica 107, 137-152.\vspace{-6pt}

\item  Vassilev, V., 1988. Group properties of a class of fourth-order
partial differential equations. Annals of the University of Sofia 82 (Part
II - Mechanics), 163-178.\vspace{-6pt}

\item  Vassilev, V., 1991. Group analysis of a class of equations of the
plate and shell theory. Ph.D. thesis, Institute of Mechanics and
Biomechanics, Bulgarian Academy of Sciences, Sofia.\vspace{-6pt}

\item  Vassilev, V., 1997. Application of Lie groups to the theory of shells
and rods. Nonlinear Analysis 30 (8), 4839-4848.\vspace{-6pt}

\item  Washizu, K., 1982. Variational Methods in Elasticity and Plasticity.
Pergamon Press, Oxford.
\end{description}

\end{document}